\documentclass[twocolumn]{autart}
\usepackage[usenames,dvipsnames,svgnames,table]{xcolor}
\usepackage{amssymb,latexsym,amsfonts,amsmath,bbm}
\usepackage[cal=cm,bb=pazo]{mathalfa}
\usepackage{mathrsfs}
\usepackage{dsfont}
\usepackage{mathtools}

\usepackage{wrapfig} % for wrapping text around figures

\newenvironment{authorbio}[2][]{%
	\par\addvspace{1em}%
	\noindent
	\if\relax\detokenize{#1}\relax % check if photo is empty
	\else
	\begin{wrapfigure}{l}{0.8in} % narrower than image width
		\vspace{-5pt} % tweak vertical alignment if needed
		\includegraphics[width=1in,height=1.25in,clip,keepaspectratio]{#1}
	\end{wrapfigure}%
	\fi
	\noindent\textbf{#2} % NAME
}{\par\addvspace{2em}}

\usepackage{graphicx}
\usepackage{eso-pic}
\usepackage{epsfig}
\usepackage{mathtools}
\usepackage{tikz}
\usepackage{circuitikz}
\usetikzlibrary{positioning,calc}
\usetikzlibrary{decorations.markings,shapes,arrows}
\usetikzlibrary{automata}
\usetikzlibrary{shapes}
\usetikzlibrary{calc,shapes,arrows}
\usepackage{mdwlist}
\usepackage{refcount}
\usepackage{comment}
\usepackage{ragged2e}
\usepackage{xcolor}
\usepackage{graphicx}
\usepackage{subfig}
\usepackage{natbib}
\usepackage{xcolor}
\definecolor{frenchblue}{rgb}{0.0, 0.45, 0.73}
\usepackage{url}
\newcommand{\R}{{\mathbb{R}}}
\newcommand{\Rpz}{{\mathbb{R}_{\geq 0}}}
\newcommand{\Rp}{{\mathbb{R}_{> 0}}}
\newcommand{\N}{{\mathbb{N}}}
\newcommand{\Np}{{\mathbb{N}_{\geq 1}}}
\newcommand{\I}{{\mathbb{I}}}
\newcommand{\V}{{\mathbb{V}}}
\newcommand{\In}{{\mathbfcal{I}}}
\newcommand{\Oi}{{\mathbfcal{O}}}
\newcommand{\Opo}{{\hat{\mathbfcal{O}}^+_1}}
\newcommand{\Opt}{{\hat{\mathbfcal{O}}^+_2}}
\newcommand{\D}{{\mathbfcal{D}}}

\newcommand{\Opoo}{{\mathbfcal{O}^+_1}}

\newcommand{\avec}{{\mathrm{a}}}
\newcounter{corollary}

\newtheorem{theorem}{Theorem}   
\newtheorem{lemma}[theorem]{Lemma}
\newtheorem{problem}{Problem}

\newtheorem{remark}{Remark}
\newtheorem{definition}{Definition}
\newtheorem{assumption}{Assumption}
\newenvironment{proof}{\textbf{Proof:}}{\hfill$\blacksquare$}
\usepackage[many]{tcolorbox}
\usetikzlibrary{calc}
\tcbuselibrary{skins}

\usepackage{algorithm}
\usepackage{algorithmic}

\newtcolorbox{resp}[1][]{%
	enhanced jigsaw,%
	colback=gray!5!white,%
	colframe=gray!80!black,%
	size=small,%
	boxrule=1pt,%
	halign title=flush center,%
	coltitle=black,%
	breakable,%
	drop shadow=black!50!white,%
	attach boxed title to top left={xshift=1cm,yshift=-\tcboxedtitleheight/2,yshifttext=-\tcboxedtitleheight/2},%
	minipage boxed title=3cm,%
	boxed title style={%
		colback=white,%
		size=fbox,%
		boxrule=1pt,%
		boxsep=2pt,%
		underlay={%
			\coordinate (dotA) at ($(interior.west) + (-0.5pt,0)$);
			\coordinate (dotB) at ($(interior.east) + (0.5pt,0)$);
			\begin{scope}[gray!80!black]
				\fill (dotA) circle (2pt);
				\fill (dotB) circle (2pt);
			\end{scope}
		}%
	},%
	#1%
}
\makeatother

\allowdisplaybreaks

\begin{document}
	
	\begin{frontmatter}
		
		\title{Data-Driven Adaptive Second-Order Sliding Mode Control with Noisy Data\thanksref{footnoteinfo}} 
		
		\thanks[footnoteinfo]{This paper was not presented at any IFAC conference. Corresponding author: Behrad Samari.}
		
		\author[Newcastle]{Behrad Samari}\ead{b.samari2@newcastle.ac.uk},    
		\author[Milan]{Gian Paolo Incremona}\ead{gianpaolo.incremona@polimi.it},               
		\author[Pavia]{Antonella Ferrara}\ead{antonella.ferrara@unipv.it}, 
		\author[Newcastle]{Abolfazl Lavaei}\ead{abolfazl.lavaei@newcastle.ac.uk}
		
		\address[Newcastle]{School of Computing, Newcastle University, United Kingdom}  
		\address[Milan]{Dipartimento di Elettronica, Informazione e Bioingegneria, Politecnico di Milano, Italy}             
		\address[Pavia]{Dipartimento di Ingegneria Industriale e dell'Informazione,
			University of Pavia, Italy}

		
		\begin{abstract}                          
			This paper proposes a data-driven approach to designing adaptive suboptimal second-order sliding mode (ASSOSM) controllers for a class of single-input nonlinear systems with partially unknown dynamics, subject to both matched and unmatched disturbances. {We first view the system as comprising two coupled dynamics,} referred to as the upper and lower dynamics, {with the last state serving as a virtual input to the upper dynamics. The proposed control-design methodology then follows a two-stage procedure: \emph{(i)} designing a virtual state-feedback control law for the upper dynamics and \emph{(ii)} synthesizing an ASSOSM controller for the full-order system.} To this end, we collect noise-corrupted data from the system throughout a finite-time experiment. We then formulate a data-dependent condition, whose feasibility enables the design of a virtual state-feedback control law that renders the closed-loop upper dynamics input-to-state stable with respect to the unmatched disturbance. Building on this virtual state-feedback control law, we subsequently propose a data-driven nonlinear sliding variable, based on which an ASSOSM controller is designed for the full-order system. The state trajectories of the resulting closed-loop system are semiglobally ultimately bounded (S-GUB), with the ultimate bound explicitly depending on the magnitude of the unmatched disturbance. In particular, the control design parameters can be selected for any prescribed bounded set of initial conditions so that the state trajectories of the closed-loop system are S-GUB. Moreover, the effect of the matched disturbance is totally rejected after a finite time. The effectiveness of the proposed method is satisfactorily demonstrated in the simulation.
		\end{abstract}
		
	\end{frontmatter}
	
	\section{Introduction}\label{sec:intro}
	Controlling nonlinear systems in uncertain environments has long been a central challenge in the control community, inspiring the development of diverse strategies to {handle} uncertainties. Among these, sliding mode (SM) control has {received} significant attention {due to} its robustness against various {classes of} uncertainties. Specifically, SM controllers drive the system state trajectories to a predefined surface, known as the \emph{sliding manifold}, in finite time  {and ensure} that they remain confined to it thereafter. Notably,  {the} system behavior on the sliding manifold  {is} unaffected by \emph{matched} uncertainties, \emph{i.e.}, uncertainties that enter through the same channel as the control input. Consequently, SM control design involves two main steps: \emph{(i)} constructing an appropriate sliding variable,  {and hence the corresponding sliding manifold,} to achieve the desired dynamic performance, and \emph{(ii)} designing a control law,  {typically} discontinuous on the sliding manifold, that ensures the state trajectories reach the manifold in finite time and remain on it thereafter~\citep{ferrara2019advanced}.
	
	Despite their advantages, SM controllers are prone to \emph{chattering} in practical implementations, \emph{i.e.}, high-frequency oscillations in the  {control input} caused by the finite switching rate of the control law, whereas ideal implementations assume switching at infinite frequency~\citep{boiko2007analysis,levant2010chattering,utkin2015discussion}.  To mitigate this phenomenon, higher-order sliding mode (HOSM) controllers have been introduced to shift the discontinuity, which is essential for finite-time convergence to the sliding manifold, to a higher-order time derivative of the control input, thereby allowing a continuous control input to be applied to the system~\citep{levant2003higher,edwards2016adaptive}. Among HOSM controllers, \emph{second-order} sliding mode (SOSM) controllers are particularly  {appealing due to} their reduced complexity,  {which makes} them more practical for implementation~\citep{bartolini1998chattering,incremona2016adaptive,ding2018new}.
	
	However, the aforementioned studies generally assume the availability of system-model knowledge to facilitate the design of the sliding variable, an assumption that is often unrealistic in practical applications. To overcome this critical challenge, two distinct yet valuable approaches have emerged: indirect and direct data-driven methods. The \emph{indirect} approach employs system identification to construct a model, which is then used for  {control design in a model-based manner.} Yet, accurately identifying complex nonlinear systems can be computationally demanding or even infeasible~\citep{kerschen2006past,hou2013model}. In contrast, \emph{direct} data-driven methods, including the approach proposed in this paper, bypass the identification phase by using the collected data directly for control design. This eliminates the two-step process inherent to indirect methods, thereby offering a more streamlined solution~\citep{dorfler2022bridging}.
	
	Motivated by the advantages of direct data-driven approaches, a growing body of research has focused on designing controllers directly from measured input--state data. Within this context, numerous studies have leveraged the \emph{informativity} approach introduced by~\citet{8960476}. Broadly speaking, this framework characterizes the set of systems compatible with the collected data and provides general tools for deriving necessary and sufficient conditions for data-based analysis and control design. For instance, \citet{van2023quadratic} provide necessary and sufficient conditions for data-based quadratic stabilization in the discrete-time setting, while comparable results are established by~\citet{10623295} in the continuous-time setting for linear time-invariant systems.
	While invaluable, this line of work has mainly focused on linear dynamical systems, with only a few recent studies addressing certain classes of nonlinear systems. For instance, \citet{11479884} propose a global stabilization approach for a class of polynomial systems that accommodates a broader class of Lyapunov function candidates. However, their framework is restricted to systems with polynomial dynamics and does not explicitly address nonvanishing process or actuator disturbances acting on the closed-loop system.
	
	 {Alternatively, inspired by the contribution of~\citet{de2019formulas}, which exploits persistently exciting data~\citep{willems2005note} for control design, direct data-driven control has been extensively studied for broader classes of polynomial systems~\citep{guo2021data,bisoffi2022data} and nonlinear systems with beyond-polynomial nonlinearities~\citep{de2023learning,monshizadeh2024meta,hu2025enforcing}; see the survey by~\citet{martin2023guarantees} for recent results.} Despite these advancements, only a few studies in this line of work have considered the problem of designing controllers that render the closed-loop system input-to-state stable (ISS) with respect to exogenous disturbances. In this direction, \citet{chen2025data} propose an approach for designing input-to-state stabilizing controllers for polynomial systems subject to actuator and process disturbances. However, their framework assumes exact access to state derivative data, which is restrictive in practice. Relatedly, \citet{zaker2025data} and~\citet{zaker2026data} address global stabilization of (infinite) networks of polynomial subsystems by designing subsystem-level input-to-state stabilizing controllers with respect to internal disturbances. Nonetheless, the internal disturbances are measurable, as they correspond to the states of other subsystems, and therefore differ fundamentally from the nonvanishing, unmeasurable exogenous disturbances considered in this paper.
	
	Recently, a few data-driven SM control frameworks have been proposed to enhance robustness against disturbances and uncertainties. In this regard, \citet{lan2024data} offer an SM control design for partially unknown nonlinear systems, while \citet{riva2024data} and~\citet{samari2025data} present \emph{integral} SM controllers for linear systems and interconnected networks of unknown nonlinear subsystems, respectively. While promising, these approaches  {remain susceptible to chattering, which may hinder their practical implementation.} In addition, \citet{samari2025data} assume that data are collected under disturbance-free conditions, a requirement that is not imposed in our setting.
	
	\textbf{Central Contribution.} The main objective of this work is to develop a data-driven approach for designing adaptive suboptimal second-order sliding mode (ASSOSM) controllers for a class of single-input nonlinear systems with partially unknown dynamics, subject to both matched and unmatched disturbances, such that the state trajectories of the resulting closed-loop system are semiglobally ultimately bounded (S-GUB). The primary contributions of the paper are stated hereafter:
	
	\begin{enumerate}
		\item[\textit{(i)}] Unlike conventional SOSM control approaches (\emph{e.g.}, the work by~\citet{bartolini1998chattering}),  {which} rely on system knowledge to construct the sliding variable,  {the proposed framework does not require an exact system model.} More importantly, in contrast to the linear sliding variables commonly adopted in the literature~\citep{bartolini1998chattering}, which are generally insufficient for the objective of this paper, we propose a \emph{nonlinear} sliding variable systematically designed from data.
		\item[\textit{(ii)}] By virtue of generating SOSMs, the proposed data-driven ASSOSM controller mitigates chattering, which is beneficial for practical applications. Moreover, when only nonvanishing actuator disturbances are present, their effect is rejected after a finite time, and the closed-loop system is rendered semiglobally asymptotically stable (S-GAS) at the origin. In contrast, the framework by~\citet{chen2025data} guarantees that the closed-loop system is ISS with respect to these disturbances. Furthermore, while the aforementioned framework is specifically developed for polynomial systems, our approach is applicable to systems with beyond-polynomial dynamics.
		\item[\textit{(iii)}] The proposed framework simultaneously accounts for three distinct sources of noise/uncertainty affecting the collected data: \emph{(a)} noise-corrupted state derivative data, \emph{(b)} unknown disturbances acting during data collection, and \emph{(c)} noisy input data, for which neither a noise bound nor its distribution is required. This differs from previous studies in which such information explicitly appears in the analysis; \emph{e.g.}, see~\citep[Remark~8]{guo2021data}.
		\item[\textit{(iv)}] For the considered system class, the proposed framework requires no predefined library of functions to represent the nonlinear, not necessarily polynomial, functions in the lower dynamics.
	\end{enumerate}
	
	\textbf{Notation.} The sets of real, non-negative real, and positive real numbers are denoted by \(\R\), \(\Rpz\), and \(\Rp\), respectively. The sets of non-negative and positive integers are denoted by \(\N\) and \(\Np\), respectively. The identity matrix of dimension \(n\times n\) is denoted by \(\I_n\), while \(\mathbf{0}_n\) and \(\mathbf{0}_{n\times m}\) denote the zero vector in \(\R^n\) and the zero matrix in \(\R^{n\times m}\), respectively. For vectors \(x_i\in\R^n\), \(i\in\{1,\ldots,N\}\), their horizontal concatenation is written as \(x=[x_1\ \ldots\ x_N]\in\R^{n\times N}\). For a symmetric matrix \(P\), \(P\succ0\) (\(P\succeq0\)) denotes that \(P\) is positive definite (positive semi-definite), while \(P\prec0\) (\(P\preceq0\)) implies that \(P\) is negative definite (negative semi-definite). Given \(P\succ0\), $\sqrt{P}$ denotes the (unique) positive-definite matrix $\bar P$ such that $\bar P^2 = P$. The minimum and maximum eigenvalues of a symmetric matrix \(P\) are denoted by \(\lambda_{\min}(P)\) and \(\lambda_{\max}(P)\), respectively. The Euclidean norm of \(x\in\R^n\) is denoted by \(\Vert x\Vert\), and \(|y|\) denotes the absolute value of \(y\in\R\). In a symmetric matrix, the symbol \(\star\) denotes the transpose of the corresponding off-diagonal block.
	
	\section{Problem Formulation}\label{sec:problem}
	
	\subsection{System Description}
	 {We first introduce the system class of interest, which is inspired by that considered by~\citet{zhang2000adaptive}.}
	
	\begin{definition}\label{def:sys}
		A continuous-time perturbed nonlinear system (\textsc{ct-PNS}) is described by\footnote{ {Throughout the paper, time arguments are omitted whenever no confusion arises; for instance, \(x_i\), \(u\), and \(d_i\) in the system dynamics stand for \(x_i(t)\), \(u(t)\), and \(d_i(t)\), respectively.}}
		\begin{align}
				\Sigma\!: \begin{cases}
					\dot{x}_i = p_i(x_1, x_2, \dots, x_{n-1})  + b_i x_{n} + d_i, \; 1 \!  \leq \! i \! \leq \! n \! - \! 1,\\
					\! \dot{x}_n = f(x) + b_n u + d_n,
				\end{cases}\label{eq:sys}
		\end{align}
		where $x \coloneq [x_1 \; x_2 \; \ldots \; x_n]^{\top} \in \R^n$ is the state vector, $u \in \R$ is the control input, and $d \coloneq [d_1 \; d_2 \; \ldots \; d_n]^\top \in \R^n$ is the disturbance vector, with $d_i \in \R$ denoting the $i$-th disturbance component for all $i \in \{1,2,\ldots,n\}$.
		Additionally, for each $i \in \{1,2,\ldots,n-1\}$, $p_i:\R^{n-1} \to \R$ is an unknown polynomial function satisfying $p_i(\mathbf{0}_{n-1})=0$, and $f:\R^n \to \R$ is an unknown (possibly highly nonlinear) continuously differentiable function satisfying $f(\mathbf{0}_n)=0$. Moreover, the constants $b_i \in \R$ for all $i \in \{1,2,\ldots,n-1\}$ and $b_n \in \Rp$ are assumed to be unknown.
		Since each $p_i$ is polynomial and $f$ is continuously differentiable, the right-hand side of~\eqref{eq:sys} is locally Lipschitz in $x$.
		Hence, for every initial condition $x_0 \coloneq x(0) \in \R^n$, every locally essentially bounded input $u(\cdot)$, and every locally essentially bounded disturbance $d(\cdot)$, system~\eqref{eq:sys} admits a unique state trajectory on a nonzero time interval.
		\hfill$\square$
	\end{definition}
	
	While the exact structure of each polynomial function $p_i$, $i \in \{1,\ldots,n-1\}$, is unknown, we assume that an upper bound on the maximum degree of these polynomial functions is available. Moreover, as stated in Definition~\ref{def:sys}, the unknown function $f$ may comprise a broad range of nonlinear functions, provided that it is continuously differentiable; hence, $f$ is not restricted to polynomial functions. To handle such nonlinear dynamics, existing data-driven approaches typically assume the availability of a library of functions capable of representing $f$~\citep{de2023learning,hu2025enforcing,monshizadeh2024meta}.
	In contrast, the approach proposed in this work does not rely on this assumption, thereby reducing the knowledge required about the system. For the sake of fairness, however, it should be acknowledged that the previously cited studies can handle more general classes of nonlinear systems. Nevertheless, many classical dynamical systems, including tunnel diode circuits, Van der Pol oscillators, and inverted pendulums, among others, take the form of the \textsc{ct-PNS} introduced in~\eqref{eq:sys}~\citep{khalil2002nonlinear}.
	
	Subsequently, one can reformulate the \textsc{ct-PNS}~\eqref{eq:sys} as
		\begin{align}
			\Sigma  :  \begin{cases}
				\dot{x}_r = \mathcal{A} \mathcal{M}(x_r) + \avec x_n + d_r,\\
				\hspace{-0.03cm}\dot{x}_n = f(x) + b_n u + d_n,
			\end{cases}\label{eq:sys_reformulated1}
		\end{align}
		where $x_r \coloneq [x_1 \; x_2 \; \dots \; x_{n-1}]^{\top} \in \R^{n-1}$, $d_r \coloneq [d_1 \; d_2 \; \ldots$ $d_{n - 1}]^\top \in \R^{n - 1}$ is the \emph{unmatched} disturbance vector, and $\mathcal{M} : \R^{n-1} \rightarrow \R^N$, with $\mathcal{M}(\mathbf{0}_{n-1}) = \mathbf{0}_N$, denotes the library of non-constant monomials constructed from the available degree bound, containing all monomials in $x_r$ whose degrees are at least one and do not exceed this bound.
		Additionally, $\mathcal{A} \in \R^{(n - 1) \times N}$ is an unknown matrix comprising the unknown coefficients of the polynomial functions, while $\avec \coloneq [b_1 \; b_2 \; \ldots \; b_{n-1}]^{\top} \in \R^{n - 1}$ is an unknown vector. Without loss of generality, since $\mathcal{M}(\mathbf{0}_{n-1})=\mathbf{0}_N$ and each entry of $\mathcal{M}$ is a monomial in $x_r$ of degree at least one, $\mathcal{M}$ can be factorized as
		\(
		\mathcal{M}(x_r)=\Omega(x_r)x_r
		\)
		for all $x_r \in \R^{n-1}$,
		where $\Omega:\R^{n-1}\to\R^{N\times (n-1)}$ is a matrix-valued polynomial function whose entries are monomials (possibly including constants) in $x_r$. Accordingly, one can rewrite the \textsc{ct-PNS}~\eqref{eq:sys_reformulated1} as
		\begin{align}
			\Sigma  :  \begin{cases}
				\dot{x}_r = \mathcal{A} \Omega(x_r)x_r + \avec x_n + d_r,\\
				\hspace{-0.03cm}\dot{x}_n = f(x) + b_n u + d_n.
			\end{cases}\label{eq:sys_reformulated}
		\end{align}
		We note that, while both $d_r$ and $d_n$ are exogenous inputs, the unmatched disturbance $d_r$ represents process disturbances affecting the dynamics outside the control channel, whereas the matched disturbance $d_n$ captures disturbances entering through the same channel as the control input and can therefore account for both process and actuator disturbances.
		
		\begin{remark}
			Assuming that an upper bound on the polynomial degree is known, consistent with the prior literature, does not impose a significant restriction. In many practical scenarios, including electrical and mechanical systems, such structural information can typically be derived from first principles~\citep{guo2021data}. Moreover, although the function $f$ may comprise arbitrary continuously differentiable functions, our method does not rely on any prespecified function library to represent it, thereby relaxing a commonly adopted assumption in earlier studies~\citep{de2023learning,hu2025enforcing,monshizadeh2024meta}. \hfill$\square$
		\end{remark}
		
		We now introduce the following assumption on the disturbances $d_r$ and $d_n$, which is commonly adopted in the SM literature~\citep{ferrara2019advanced}.
		
		\begin{assumption}\label{assump:bound}
			The unmatched and matched disturbances $d_r$ and $d_n$ are continuously differentiable and, for all $t \in \Rpz$, satisfy $\|d_r(t)\| \leq \bar{d}_1$, $\|\dot{d}_r(t)\| \leq \bar{d}_2$, $|d_n(t)| \leq \bar{d}_3$, and $|\dot{d}_n(t)| \leq \bar{d}_4$, where $\bar{d}_1 \in \Rp$ is a known constant, while $\bar{d}_2,\bar{d}_3,\bar{d}_4 \in \Rp$ are all unknown constants. \hfill$\square$
		\end{assumption}
		
		As is evident from Assumption~\ref{assump:bound}, the proposed framework only requires the \emph{existence} of the constants $\bar d_2$ through $\bar d_4$; neither their numerical values nor any estimates thereof are needed. Moreover, in practical applications, $\bar d_1$ can be selected based on physical insight into the source of the unmatched disturbance, such as wind disturbances, friction uncertainty, modeling residuals, or parameter variations. A conservative estimate of the maximum admissible magnitude of these effects can therefore be used.
		
		With the \textsc{ct-PNS}~$\Sigma$ and the associated assumptions now fully described, the following subsection  {presents the main objective of the paper, outlines the proposed solution strategy, and formally states the primary problem addressed in this work.}
		
		\subsection{Design Approach}
		The aim of this paper is to design an ASSOSM controller for the \textsc{ct-PNS}~\eqref{eq:sys_reformulated} without exact model knowledge, such that the state trajectories of the resulting closed-loop system are S-GUB, with an ultimate bound that is independent of the magnitude of $d_n$ while explicitly depending on that of $d_r$. As a direct consequence, in the absence of the unmatched disturbance $d_r$, the closed-loop system is S-GAS at $x = \mathbf{0}_n$. To address this objective, we propose a two-step, purely data-driven control design methodology. More precisely, in the first step, we treat \( x_n \) as a virtual input for the \( x_r \)-dynamics\footnote{Throughout the paper, we interchangeably use the terms ``$x_r$-dynamics" and ``upper dynamics" to refer to the first dynamics given in~\eqref{eq:sys_reformulated}.} of~\eqref{eq:sys_reformulated} and design it as a virtual feedback law (denoted by $\varphi$) that renders the closed-loop upper dynamics ISS via a Lyapunov-based argument. For completeness, in the subsequent definition, we formally recall the notion of ISS Lyapunov functions, whose existence certifies that the closed-loop system is ISS, referring to the upper dynamics of~\eqref{eq:sys_reformulated}.
		
		\begin{definition}\label{def:ISS}
			Consider the upper dynamics of the \textsc{ct-PNS}~\eqref{eq:sys_reformulated}, and let $x_n \coloneq \varphi(x_r)$, where $\varphi:\R^{n-1}\to\R$ is a polynomial function satisfying $\varphi(\mathbf{0}_{n-1})  =  0$. A continuously differentiable function $\V  :  \R^{n-1} \to \Rpz$ is said to be an ISS Lyapunov function for the $x_r$-dynamics in~\eqref{eq:sys_reformulated} if there exist constants $\underline{\alpha}, \, \overline{\alpha}, \, \kappa \in \Rp$ and $\rho \in \Rpz$ such that
				\begin{subequations}\label{eq:ISS_Lyap}
				\begin{itemize}
					\item $\forall x_r \in \R^{n - 1} \! :$
						\begin{align}
							\underline{\alpha} \Vert x_r \Vert^2 \leq \V(x_r) \leq \overline{\alpha} \Vert x_r \Vert^2 , \label{eq:ISS1}
						\end{align}
						\item $\forall x_r \in \R^{n - 1}$ and $\forall d_r \in \R^{n - 1} \! :$
						\begin{align}
							\dot{\V}(x_r) \leq -\kappa \V(x_r) + \rho \Vert d_r \Vert^2, \label{eq:ISS2}
						\end{align}
					where $\dot{\V}$ denotes the derivative of $\V$ along the trajectories of the $x_r$-dynamics in~\eqref{eq:sys_reformulated}, given by
					\begin{align*}
						\dot{\V}(x_r) =  \frac{\partial \V}{\partial x_r}(\mathcal{A} \Omega(x_r)x_r + \avec \varphi(x_r) + d_r).\tag*{$\text{\textcolor{black}{$\square$}}$}
					\end{align*}
			\end{itemize}
		\end{subequations}
		\end{definition}
		
		To proceed with the proposed two-step design, while ensuring that the virtual control problem for the upper dynamics is well posed, we introduce the following assumption.
		
		\begin{assumption}\label{assump:varphi}
			There exists a polynomial function $\varphi:\R^{n-1}\to\R$, with $\varphi(\mathbf{0}_{n-1})=0$, such that the upper dynamics of~\eqref{eq:sys_reformulated} under $x_n=\varphi(x_r)$ admit an ISS Lyapunov function in the sense of Definition~\ref{def:ISS}. \hfill$\square$
		\end{assumption}
		
		In the second step of our proposed approach, based on \(\varphi\) designed in the initial step, we construct the sliding variable and design the ASSOSM controller, ensuring that the state trajectories of the resulting closed-loop system are S-GUB and the effect of $d_n$ on the state dynamics is rejected after a finite time. More concretely, we design the sliding variable $\sigma : \R^n \rightarrow \R$ as
		\begin{align}
			\sigma = x_n - \varphi(x_r). \label{eq:sliding_manifold}
		\end{align}
		At the same time, under the designed ASSOSM controller, both \( \sigma \) and \( \dot{\sigma} \) are driven to zero in finite time, which, considering~\eqref{eq:sliding_manifold}, implies that \( x_n = \varphi(x_r) \) after a finite time. In other words, for any initial condition in an arbitrarily large prespecified bounded set, the parameters of the ASSOSM controller can be chosen to drive the system trajectories to the sliding manifold $\sigma = 0$ in finite time and keep them on it thereafter.
	
	As a consequence of the chosen sliding variable and the design procedure, one can deduce that, on the sliding manifold $\sigma = 0$, the behavior of $x_n$ is essentially inherited from that of $x_r$.
	Accordingly, since $\varphi$ renders the $x_r$-dynamics in~\eqref{eq:sys_reformulated} ISS with respect to $d_r$, $x_r$ remains bounded for any bounded $d_r$. This implies that $\varphi$ also remains bounded; consequently, when $\sigma = 0$, $x_n$ is likewise bounded (cf.~\eqref{eq:sliding_manifold}). We also note that, during the finite time interval in which $\sigma \neq 0$, $\varphi$ renders the $x_r$-dynamics in~\eqref{eq:sys_reformulated} ISS with respect to both $d_r$ and $\sigma$. With these observations in place and recalling that the ASSOSM controller drives trajectories from any initial condition within an arbitrarily large bounded set to the sliding manifold $\sigma = 0$ in finite time and maintains them there thereafter, it follows that the state trajectories of the resulting full-order closed-loop system are S-GUB under the control input $u$. Moreover, in the absence of $d_r$, the closed-loop system is S-GAS at $x=\mathbf{0}_n$.
	
	 {We note that, although \(\varphi\) renders the upper dynamics ISS (a global property), the proposed ASSOSM controller provides semiglobal guarantees, as it requires the existence of a finite bound on a term appearing in the second time derivative of the sliding variable, which depends on the state variables.} If the state variables are unbounded, such a bound does not exist, and therefore, the ASSOSM framework cannot be applied. We emphasize that the key motivation for employing ASSOSM control rather than its classical, non-adaptive counterpart~\citep{bartolini1998chattering} is its suitability for data-driven settings, where the aforementioned bound and the constants $\bar d_2$ through $\bar d_4$ are assumed to be unknown (cf. Assumption~\ref{assump:bound}).
	Moreover, in the partially unknown setting considered, $\varphi$ and, consequently, the sliding variable in~\eqref{eq:sliding_manifold} should be designed based on the collected data. With these critical challenges in mind, we now formally state the main problem addressed in this work.
	
	\begin{resp}
		\begin{problem}\label{problem1}
			Consider the  {partially unknown} \textsc{ct-PNS}~\eqref{eq:sys_reformulated}. By collecting  {noise-corrupted} data from  {a single trajectory of the system}, design a continuous control input that  {steers the system trajectories to the sliding manifold $\sigma = 0$, with $\sigma$ as in~\eqref{eq:sliding_manifold},} thereby ensuring that the state trajectories of the closed-loop system are S-GUB. \hfill$\square$
		\end{problem}
	\end{resp}
	
	In the following section, we present our data-driven framework addressing Problem~\ref{problem1}.
	
	\section{Data-Driven Framework}\label{sec:solution}
	In this section, we introduce our data-driven ASSOSM control design. In particular, Section~\ref{subsec:manifold} outlines the noisy data collection process and shows how the collected data can be exploited to design $\varphi$ rendering the closed-loop $x_r$-dynamics of the \textsc{ct-PNS}~\eqref{eq:sys_reformulated} ISS. Section~\ref{subsec:controller} presents the ASSOSM approach for designing the control input $u$, which drives both $\sigma$ and $\dot{\sigma}$ to zero in finite time, thereby ensuring that the state trajectories of the full-order closed-loop system are S-GUB.
	
	\subsection{Data-Driven Sliding Variable Design}\label{subsec:manifold}
	To design $\varphi$ in a data-driven manner,  {we collect input--state data from the system over the time interval $[t_0, \,  t_0+(\mathcal{T}- 1)\tau]$, where $\mathcal{T} \in \Np$ is the number of sampled data points and $\tau \in \Rp$ denotes the sampling time,} whose choice is arbitrary. In fact, the proposed result remains valid for unevenly spaced sampled measurements; however, for notational simplicity, the construction is presented using uniformly sampled data. Subsequently, the data matrices containing the collected input--state data are formed as
	\begin{subequations}\label{eq:IS-DATA}
		\begin{align}
			\In & \coloneq \big[u(t_0) \;\; u(t_0 \!+\! \tau) \;\; \dots \;\; u(t_0\!+\!(\mathcal{T}\!-\! 1)\tau)\big]\!, \label{eq:data-Input}\\
			\Oi_1 & \coloneq \big[x_r(t_0) \;\; x_r(t_0 \!+\! \tau) \;\; \dots \;\; x_r(t_0\!+\!(\mathcal{T}\!-\! 1)\tau)\big] \!,\label{eq:data:state}\\
			\Oi_2 & \coloneq \big[x_n(t_0) \;\; x_n(t_0 \!+\! \tau) \;\; \dots \;\; x_n(t_0\!+\!(\mathcal{T}\!-\! 1)\tau)\big] \!. \label{eq:data_virtual}
		\end{align}
		Moreover, since $\mathcal{M}$ is a known library of monomials, we also construct, using the data in~\eqref{eq:data:state}, the data matrix
		\begin{align}
			\mathbfcal{J} \! \coloneq \! \! \big[\mathcal{M}(x_r(t_0)\!) \: \mathcal{M}(x_r(t_0 \!+\! \tau)\!) \: \dots \: \mathcal{M}(x_r(t_0\!+\!(\mathcal{T}\!-\! 1)\tau)\!)\!\big] \!, \label{eq:dictionary_data}
		\end{align}
		which is utilized in the following analysis.
		We note that, while the state measurements in~\eqref{eq:data:state} and~\eqref{eq:data_virtual} are assumed to be noise-free, as is common in the data-driven nonlinear control literature~\citep{guo2021data,11479884}, the proposed framework naturally accommodates measurement noise in the input data~\eqref{eq:data-Input} (\emph{i.e.}, when the input stored in the dataset differs from the true input), without requiring any structural assumptions on the noise. This feature of the proposed framework is particularly appealing from a practical perspective, especially compared with much of the existing literature, where the input data in~\eqref{eq:data-Input} is typically assumed to be available precisely despite the presence of input measurement noise in real-world applications.
	\end{subequations}
	
	As the \textsc{ct-PNS}~\eqref{eq:sys_reformulated} is subject to process and actuator disturbances (\emph{i.e.}, unmatched and matched disturbances), denoted by $d_r$ and $d_n$, respectively, the collected data are naturally affected by these disturbances. We introduce the following data matrices relevant to the disturbances:
		\begin{subequations}\label{eq:D-DATA}
			\begin{align}
				\D_1 & \coloneq \big[d_r(t_0) \;\; d_r(t_0 \!+\! \tau) \;\; \dots \;\; d_r(t_0\!+\!(\mathcal{T}\!-\! 1)\tau)\big]\!, \label{eq:data-dist}\\
				\D_2 & \coloneq \big[d_n(t_0) \;\; d_n(t_0 \!+\! \tau) \;\; \dots \;\; d_n(t_0\!+\!(\mathcal{T}\!-\! 1)\tau)\big]\!. \label{eq:data-dist2}
			\end{align}
	\end{subequations}
	Notice that both $\D_1$ and $\D_2$ are unknown and inaccessible, in line with real-world scenarios; they are introduced here solely to facilitate the subsequent analysis. Nevertheless, under Assumption~\ref{assump:bound}, we know that $\Vert d_{r_j} \Vert \leq \bar d_1$ for all $j \in \{1, \ldots, \mathcal T\}$, where $d_{r_j}$ denotes the $j$-th column of $\D_1$. Accordingly, one has
	\begin{align}
		\D_1 \D_1^\top = \sum_{j = 1}^{\mathcal{T}} d_{r_j} d_{r_j}^\top \preceq \sum_{j = 1}^{\mathcal{T}} \Vert d_{r_j} \Vert^2 \I_{n - 1} \preceq \mathds{D}, \label{eq:bound_dr}
	\end{align}
	where $\mathds{D} \coloneq \bar{d}_1^2 \mathcal{T} \I_{n - 1}$.
	We note that such an analysis is not required within our framework for $\D_2$, as it inherently provides robustness guarantees against the matched disturbance $d_n$ (analogously to what happens in the case of noisy measurements in $\In$).
	
	To capture the system behavior using data, we also require the state derivative data, whose corresponding data matrices are arranged as follows:
		\begin{subequations}\label{eq:DE-DATA}
			\begin{align}
				\Opo & \coloneq \! \big[\dot{x}_r(t_0) \;\; \dot{x}_r(t_0 \!+\! \tau) \:\: \dots \:\: \dot{x}_r(t_0\!+\!(\mathcal{T}\!-\! 1)\tau)\big] \!, \label{eq:data-der-state}\\
				\Opt & \coloneq \! \big[\dot{x}_n(t_0) \;\; \dot{x}_n(t_0 \!+\! \tau) \,\, \dots \,\, \dot{x}_n(t_0\!+\!(\mathcal{T}\!-\! 1)\tau)\big] \!.\label{eq:data-xn-dot}
			\end{align}
	\end{subequations}
    As $\Opo$ and $\Opt$ cannot usually be measured directly, and since the subsequent analysis does not rely on $\Opt$ in~\eqref{eq:data-xn-dot}, we approximate only the elements of $\Opo$ as
	\begin{align}
		\dot{x}_i\left(t_0 \! +\! k\tau\right)\! = \! \frac{x_i\left(t_{0} \! + \! (k \! + \! 1)\tau\right) \! - \! x_i\left(t_0 \! + \! k\tau\right)}{\tau} \! + \! \psi_i\left(t_0 \! + \! k\tau\right)\!, \label{eq:FD}
	\end{align}
	with $i \in \{1, \ldots, n-1\}$ and $k \in \{0, 1, \dots, \mathcal T-1\}$\footnote{To approximate $\dot{x}_i(t_0+(\mathcal T-1)\tau)$ for all $i\in\{1,\ldots,n-1\}$ using~\eqref{eq:FD}, we additionally require $x_i(t_0+\mathcal T\tau)$; however, this sample is used solely for the approximation and is not included in the data matrices in~\eqref{eq:IS-DATA}.}, where the approximation error $\psi_i\left(t_0+k\tau\right)$ is proportional to $\tau$ and can be regarded as noise.
	Thus, instead of the exact $\Opo$ in~\eqref{eq:data-der-state}, we access $\Opoo$ (\emph{i.e.}, $\Opoo$ is the approximated data matrix associated with the derivative of $x_r$), where $\Opo = \Opoo + \Psi$ (cf.~\eqref{eq:FD}), with
	\[
	\Psi \coloneq \big[\psi(t_0) \;\; \psi(t_0 \!+\! \tau) \;\; \dots \;\; \psi(t_0\!+\!(\mathcal{T}\!-\! 1)\tau)\big]\!
	\]
	accounting for the approximation error. While $\Psi$ is unknown, we impose the following assumption on it~\citep{van2020noisy}.
	
	\begin{assumption}\label{assump:bound2}
		The unknown data matrix $\Psi$ satisfies
		\[
		\Psi \Psi^\top \preceq \gamma \gamma^\top,
		\]
		 {where $\gamma$ is a known matrix of appropriate dimensions}. \hfill$\square$
	\end{assumption}
	
	\begin{remark}\label{rem:assum2}
		In practice, a concrete scenario in which Assumption~\ref{assump:bound2} holds is when a constant $\bar{\psi}$ is known such that $\Vert \psi_j \Vert^2 \leq \bar{\psi}$ for all $j \in \{1, \ldots, \mathcal T\}$, where $\psi_j$ denotes the $j$-th column of $\Psi$. In such a case, by following the same steps as in~\eqref{eq:bound_dr}, Assumption~\ref{assump:bound2} is satisfied with $\gamma \coloneq \sqrt{\bar{\psi} \mathcal{T}} \I_{n-1}$, resulting in $\gamma \gamma^\top = \bar{\psi} \mathcal{T}\I_{n-1}$. \hfill$\square$
	\end{remark}
	
	\begin{remark}\label{rem:derivative-free}
		It is possible to avoid the use of forward differences in~\eqref{eq:FD} for approximating the state derivative data. Similarly to the approach proposed by~\citet[Appendix~A]{10323524}, if trajectory-level data, or sufficiently accurate integral data, are available over each sampling interval, which is typically a stronger practical requirement than having only pointwise sampled data as in~\eqref{eq:data:state} and~\eqref{eq:data_virtual}, one may alternatively integrate the $x_r$-dynamics and proceed with the subsequent analysis using sample differences of $x_r$ together with the integrals of $\mathcal M$ and $x_n$. In this case, the effect of the unmatched disturbance can be bounded over each sampling interval using Assumption~\ref{assump:bound} and incorporated into the analysis following reasoning analogous to that used in~\eqref{eq:bound_dr}. Note that when only pointwise sampled data are available, the above-mentioned integrals would need to be numerically approximated, which may introduce quadrature errors that should also be incorporated into the analysis; such errors would play a role similar to that of $\Psi$. Accordingly, in this work, we approximate state derivative data using~\eqref{eq:FD}, as it only requires pointwise sampled data, and its approximation error is explicitly captured by Assumption~\ref{assump:bound2}. For more general derivative-free data-driven formulations, primarily for linear dynamical systems, we refer the reader to the works by~\citet{10268593} and~\citet{11474915}. \hfill$\square$
	\end{remark}
	
	We now propose the following lemma, under which a closed-loop representation for the $x_r$-dynamics of the \textsc{ct-PNS}~\eqref{eq:sys_reformulated} is obtained.
	
	\begin{lemma}\label{lemma1} 
		Given the $x_r$-dynamics of the \textsc{ct-PNS}~\eqref{eq:sys_reformulated}, let $\mathcal{S} \coloneq \big[\avec \;\; \mathcal{A}\big]$. If the virtual state-feedback control law is chosen as $\varphi(x_r) = \mathbfcal{K}(x_r) \mathbfcal{P} x_r$, where $\mathbfcal{K} : \R^{n - 1} \to \R^{1 \times (n-1)}$ is a matrix-valued polynomial function and $\mathbfcal{P} \in \R^{(n  -  1) \times (n  -  1)}$ is a constant matrix, one has
		\begin{align}
			\dot{x}_r = \mathcal{S} \begin{bmatrix}
				\mathbfcal{K}(x_r) \mathbfcal{P}\\
				\Omega(x_r)
			\end{bmatrix} \! x_r + d_r \label{eq:new-closed-form}
		\end{align}
		as the resulting closed-loop $x_r$-dynamics.
		\hfill$\square$
	\end{lemma}
	
	\begin{proof}
		Considering $\mathcal{S} = \big[\avec \;\; \mathcal{A}\big]$, and under the virtual state-feedback control law $\varphi(x_r) = \mathbfcal{K}(x_r) \mathbfcal{P} x_r$, one has
		\begin{align*}
			\dot{x}_r & \hspace{-0.04cm} \overset{\eqref{eq:sys_reformulated}}{=}   \mathcal{A} \Omega(x_r)x_r + \avec x_n + d_r =  \mathcal{A} \Omega(x_r)x_r + \avec\varphi(x_r) + d_r \\
			& = (\mathcal{A}\Omega(x_r)  +  \avec  \mathbfcal{K}(x_r) \mathbfcal{P}) x_r + d_r \\
			& = \big[\avec \;\; \mathcal{A}\big] \begin{bmatrix}
				\mathbfcal{K}(x_r) \mathbfcal{P}\\
				\Omega(x_r)
			\end{bmatrix} \! x_r + d_r = \mathcal{S} \begin{bmatrix}
				\mathbfcal{K}(x_r) \mathbfcal{P}\\
				\Omega(x_r)
			\end{bmatrix} \!  x_r + d_r,
		\end{align*}
		concluding the proof.
	\end{proof}
	
	We now introduce the following theorem, which is one of the main results of this work. This theorem enables the design of $\varphi$ that renders the closed-loop $x_r$-dynamics ISS with respect to $d_r$. Consequently, this facilitates the data-driven design of the sliding variable~\eqref{eq:sliding_manifold} using  {noise-corrupted} data collected  {over} a finite-time experiment.
	
	\begin{thm}\label{thm:main}
		Given the $x_r$-dynamics of the \textsc{ct-PNS}~\eqref{eq:sys_reformulated}, let Assumptions~\ref{assump:bound}--\ref{assump:bound2} hold.
		If, for some $\kappa, \varkappa \in \Rp$, there exist a polynomial function $\varpi : \R^{n-1} \to \Rpz$, a matrix-valued polynomial function $\mathbfcal{K} : \R^{n-1} \to \R^{1 \times (n-1)}$, a constant $\epsilon \in \Rp$, and a constant matrix $\mathbfcal{Q} \in \R^{(n - 1) \times (n-1)}$, with $\mathbfcal{Q} \succ 0$, such that
			\begin{align}
				\begin{bmatrix}
					\mathbfcal{R}(x_r) & ~\begin{bmatrix}
						\mathbfcal{K}(x_r)\\
						\Omega(x_r) \mathbfcal{Q}
					\end{bmatrix}^{\!\! \top} + \varpi(x_r) \Opoo \mathbfcal{G}^\top\\
					\star & - \varpi(x_r) \mathbfcal{G}  \mathbfcal{G}^\top
				\end{bmatrix} \!\! \preceq 0, \forall x_r \in \R^{n - 1},
				\label{eq:thm}
			\end{align}
			where
			\(
			\mathbfcal{R}(x_r) \coloneq \epsilon \I_{n - 1} + \kappa \mathbfcal{Q} - \varpi(x_r) (\Opoo \Opoo^{\! \top} - \mathcal{Y}),
			\)
			with $\mathcal{Y} \coloneq (1 + \varkappa) \mathds{D} + \big(1 + \frac{1}{\varkappa}\big) \gamma \gamma^\top$,
			and
			$\mathbfcal{G} \coloneq \big[ \Oi_2^\top \;\; \mathbfcal{J}^\top \big]^\top$,
			then the virtual state-feedback control law $\varphi(x_r) = \mathbfcal{K}(x_r)\mathbfcal{P} x_r$, with $\mathbfcal{P} \coloneq \mathbfcal{Q}^{-1} \succ 0$, renders the closed-loop $x_r$-dynamics of the \textsc{ct-PNS} ISS, with the corresponding ISS Lyapunov function $\V(x_r) = x_r^\top \mathbfcal{P} x_r$, where $\underline{\alpha} \coloneq \lambda_{\min}(\mathbfcal{P})$, $\overline{\alpha} \coloneq \lambda_{\max}(\mathbfcal{P})$, and $\rho \coloneq \frac{1}{\epsilon}$. \hfill$\square$
	\end{thm}
	
	\begin{proof}
		Consider the ISS Lyapunov function $\V(x_r) = x_r^\top \mathbfcal{P} x_r$ with $\mathbfcal{P} \coloneq \mathbfcal{Q}^{-1}$, and hence, $\mathbfcal{P}^{-1} = \mathbfcal{Q}$.
		Notice that $\mathbfcal{Q} \succ 0$, and accordingly, $\mathbfcal{P} \succ 0$.
		For the candidate ISS Lyapunov function, since
			\[
			\lambda_{\min}(\mathbfcal{P}) \Vert x_r \Vert^2 \leq x_r^\top \mathbfcal{P} x_r \leq \lambda_{\max}(\mathbfcal{P}) \Vert x_r \Vert^2,
			\]
			one can conclude that condition~\eqref{eq:ISS1} is satisfied with $\underline{\alpha} \coloneq \lambda_{\min}(\mathbfcal{P})$ and $\overline{\alpha} \coloneq \lambda_{\max}(\mathbfcal{P})$.
		We now show that, under the feasibility of the proposed condition~\eqref{eq:thm}, it is guaranteed that condition~\eqref{eq:ISS2} holds. Under the virtual state-feedback control law $\varphi(x_r) = \mathbfcal{K}(x_r) \mathbfcal{P} x_r$, and according to~\eqref{eq:new-closed-form}, we have
		\begin{align*}
			\dot{\V}(x_r) & = 2x_r^\top \mathbfcal{P} \dot{x}_r \overset{\eqref{eq:new-closed-form}}{=} 2x_r^\top \mathbfcal{P} \Big( \mathcal{S} \begin{bmatrix}
				\mathbfcal{K}(x_r) \mathbfcal{P}\\
				\Omega(x_r)
			\end{bmatrix} \! x_r + d_r \Big)\\
			& = 2x_r^\top \mathbfcal{P} \mathcal{S} \begin{bmatrix}
				\mathbfcal{K}(x_r) \mathbfcal{P}\\
				\Omega(x_r)
			\end{bmatrix} \! x_r + 2x_r^\top \mathbfcal{P} d_r.
		\end{align*}
		By applying the Cauchy--Schwarz inequality~\citep{bhatia1995cauchy}, \emph{i.e.}, $x_r^\top \mathbfcal{P} d_r \le \Vert \mathbfcal{P} x_r \Vert \Vert d_r \Vert$, followed by Young's inequality~\citep{young1912classes}, \emph{i.e.}, $\Vert \mathbfcal{P} x_r \Vert \Vert d_r \Vert \le \tfrac{\epsilon}{2}\Vert \mathbfcal{P} x_r \Vert^2 + \tfrac{1}{2\epsilon}\Vert d_r \Vert^2$ for any $\epsilon \in \Rp$, one has
		\begin{align}
			& \dot{\V}(x_r) \notag\\
			& \leq 2x_r^\top \mathbfcal{P} \mathcal{S} \begin{bmatrix}
				\mathbfcal{K}(x_r) \mathbfcal{P}\\
				\Omega(x_r)
			\end{bmatrix} \! x_r + \epsilon x_r^\top \mathbfcal{P} \mathbfcal{P} x_r + \frac{1}{\epsilon} \Vert d_r \Vert^2\notag\\
			& = x_r^\top \Big( \mathbfcal{P} \mathcal{S} \begin{bmatrix}
				\mathbfcal{K}(x_r) \mathbfcal{P}\\
				\Omega(x_r)
			\end{bmatrix} + \begin{bmatrix}
				\mathbfcal{K}(x_r) \mathbfcal{P}\\
				\Omega(x_r)
			\end{bmatrix}^{\! \top} \mathcal{S}^{\top} \mathbfcal{P} + \epsilon \mathbfcal{P} \mathbfcal{P} \Big)x_r\notag\\
			& \hphantom{=} + \frac{1}{\epsilon} \Vert d_r \Vert^2\notag\\
			& = x_r^\top \mathbfcal{P} \Big( \! \mathcal{S} \! \begin{bmatrix}
				\mathbfcal{K}(x_r)\\
				\Omega(x_r) \mathbfcal{P}^{-1}
			\end{bmatrix} \! + \! \begin{bmatrix}
				\mathbfcal{K}(x_r)\\
				\Omega(x_r) \mathbfcal{P}^{-1}
			\end{bmatrix}^{\!\! \top} \!\! \mathcal{S}^{\top} \! + \! \epsilon \I_{n - 1} \! \Big) \mathbfcal{P} x_r\notag\\
			&\hphantom{=} + \frac{1}{\epsilon} \Vert d_r \Vert^2\notag\\
			& = x_r^\top \mathbfcal{P} \! \begin{bmatrix}
				\I_{n-1}\\ \mathcal{S}^\top
			\end{bmatrix}^{\!\! \top} \! \overbrace{\begin{bmatrix}
				\epsilon \I_{n - 1} & \begin{bmatrix}
					\mathbfcal{K}(x_r)\\
					\Omega(x_r) \mathbfcal{P}^{-1}
				\end{bmatrix}^{\!\! \top}\\
				\star & \mathbf{0}_{(N+1) \times (N+1)}
				\end{bmatrix}}^{\Xi_1(x_r)} \! \begin{bmatrix}
			\I_{n-1}\\ \mathcal{S}^\top
			\end{bmatrix} \mathbfcal{P} x_r\notag\\
			&\hphantom{=} + \frac{1}{\epsilon} \Vert d_r \Vert^2.\label{eq:thm1_tmp1}
		\end{align}
		At the same time, $\V(x_r) = x_r^\top \mathbfcal{P} x_r$ can be rewritten as
		\begin{align}
			\V(x_r) \! = \! x_r^\top \mathbfcal{P} \! \overbrace{\begin{bmatrix}
					\I_{n-1}\\ \mathcal{S}^\top
				\end{bmatrix}^{\!\! \top} \!\! \underbrace{\begin{bmatrix}
						\mathbfcal{P}^{-1} & \mathbf{0}_{(n - 1) \times (N + 1)}\\
						\star & \mathbf{0}_{(N+1) \times (N+1)}
				\end{bmatrix}}_{\Xi_2} \!\! \begin{bmatrix}
					\I_{n-1}\\ \mathcal{S}^\top
			\end{bmatrix}}^{\mathbfcal{P}^{-1}} \! \mathbfcal{P} x_r.\label{eq:thm1_tmp2}
		\end{align}
		For notational convenience, we define $z(x_r) \coloneq \begin{bmatrix}
			\I_{n-1}\\ \mathcal{S}^\top
		\end{bmatrix} \mathbfcal{P}x_r$.
		Considering~\eqref{eq:thm1_tmp1} and~\eqref{eq:thm1_tmp2}, if we show that
		\begin{align*}
			z^\top(x_r) \Xi_1(x_r) z(x_r) \leq -\kappa z^\top(x_r) \Xi_2 z(x_r),
		\end{align*}
		or equivalently
		\begin{align}
			z^\top(x_r) \big( \Xi_1(x_r) + \kappa \Xi_2 \big) z(x_r) \leq 0
			\label{eq:thm1_tmp3}
		\end{align}
		holds under the proposed condition~\eqref{eq:thm}, then we can conclude that condition~\eqref{eq:ISS2} is satisfied with $\rho \coloneq \frac{1}{\epsilon}$ and some $\kappa \in \Rp$. As directly enforcing $\Xi_1(x_r) + \kappa \Xi_2 \preceq 0$ is infeasible, we revisit~\eqref{eq:bound_dr} and Assumption~\ref{assump:bound2}. First, according to the $x_r$-dynamics in~\eqref{eq:sys_reformulated} and the collected data in~\eqref{eq:IS-DATA}, and recalling that $\Opo = \Opoo + \Psi$, we have
		\begin{align}
			\Opoo = \mathcal{A} \mathbfcal{J} + \avec \Oi_2 + \D_1 - \Psi = \mathcal{S} \mathbfcal{G} + \D_1 - \Psi, \label{eq:thm1_tmp4}
		\end{align}
		where $\mathbfcal{G} \coloneq \big[ \Oi_2^\top \;\; \mathbfcal{J}^\top \big]^\top$. Additionally, we know that
		\[
		(\D_1 - \Psi) (\D_1 - \Psi)^\top \preceq (1 + \varkappa) \D_1 \D_1^\top + \big(1 + \frac{1}{\varkappa}\big) \Psi \Psi^\top
		\]
		holds for any $\varkappa \in \Rp$. Consequently, considering~\eqref{eq:bound_dr} and Assumption~\ref{assump:bound2}, we have
		\begin{align}
			& (\D_1 - \Psi) (\D_1 - \Psi)^\top \preceq (1 + \varkappa) \mathds{D} + \big(1 + \frac{1}{\varkappa}\big) \gamma \gamma^\top\notag\\
			& \overset{\eqref{eq:thm1_tmp4}}{\Longrightarrow} \! (\Opoo \! - \! \mathcal{S} \mathbfcal{G}) (\Opoo \! - \! \mathcal{S} \mathbfcal{G})^\top \! \preceq \! \overbrace{(1 + \varkappa) \mathds{D} + \big(1 + \frac{1}{\varkappa}\big) \gamma \gamma^\top}^{\mathcal{Y}}\notag\\
			& \Longrightarrow \begin{bmatrix}
				\I_{n-1}\\ \mathcal{S}^\top
			\end{bmatrix}^{\!\! \top}  \overbrace{\begin{bmatrix}
					\Opoo \Opoo^{\! \top} - \mathcal{Y} & ~~-\Opoo \mathbfcal{G}^\top\\
					\star & \mathbfcal{G}\mathbfcal{G}^\top
			\end{bmatrix}}^{\Xi_3}  \begin{bmatrix}
				\I_{n-1}\\ \mathcal{S}^\top
			\end{bmatrix} \preceq 0. \label{eq:thm1_tmp5}
		\end{align}
		From~\eqref{eq:thm1_tmp5}, one gets $z^\top(x_r) \Xi_3 z(x_r) \leq 0$.
		This is precisely where the S-procedure~\citep{polik2007survey} can be used to conclude~\eqref{eq:thm1_tmp3}. More concretely, according to the S-procedure, if there exists $\varpi : \R^{n -1} \to \Rpz$ such that
		\begin{align}
			z^\top(x_r) \big( \Xi_1(x_r) + \kappa \Xi_2 -\varpi(x_r)\Xi_3 \big) z(x_r) \leq 0, \label{eq:thm1_tmp7}
		\end{align}
		then one can conclude that~\eqref{eq:thm1_tmp3} holds. Evidently, under the proposed condition~\eqref{eq:thm}, $\Xi_1(x_r) + \kappa \Xi_2 - \varpi(x_r)\Xi_3 \preceq 0$ for all $x_r \in \R^{n-1}$, meaning that~\eqref{eq:thm1_tmp7} holds. Consequently, one can deduce that~\eqref{eq:thm1_tmp3} holds; therefore
		\[
		\dot{\V}(x_r) \leq -\kappa \V(x_r) + \rho \Vert d_r \Vert^2,
		\]
		with $\rho \coloneq \frac{1}{\epsilon}$ and some $\kappa \in \Rp$, which concludes the proof.
	\end{proof}
	
	\begin{remark}\label{rem:rank_existence}
		Observe that, while condition~\eqref{eq:thm} does not require $\mathbfcal G$ to be of full row rank, such a rank condition can facilitate its feasibility. In fact, if $\mathbfcal G$ is not of full row rank, then $\mathbfcal G\mathbfcal G^\top$ is singular. In this case, condition~\eqref{eq:thm} may still hold, but its feasibility requires that each column of the block denoted by $\star$ lie, pointwise in $x_r$, in the column space of the bottom-right block; this requirement is nontrivial and can be restrictive. This potential algebraic obstruction is avoided whenever $\varpi(x_r)>0$ for all $x_r \in \R^{n-1}$ and $\mathbfcal G$ is of full row rank, since $\mathbfcal G\mathbfcal G^\top \succ 0$ and the bottom-right block is negative definite. For $\mathbfcal G$ to be of full row rank, it is necessary that $\mathcal T \geq N+1$. \hfill$\square$
	\end{remark}
	
	\begin{remark}\label{rem:non_unique}
		The solution to condition~\eqref{eq:thm} is generally non-unique and depends on the data collected during the finite-time experiment. In particular, the data matrices are affected by the initial condition, the input used during data collection, the disturbance realizations, and the sampling instants. While the proposed result does not impose an a priori restriction on the initial condition or the sampling time, these choices may influence the rank properties of $\mathbfcal G$ (cf.~Remark~\ref{rem:rank_existence}) and the derivative-approximation error in~\eqref{eq:FD}, thereby potentially affecting feasibility of condition~\eqref{eq:thm}. In fact, considering~\eqref{eq:FD}, larger values of $\tau$ may lead to greater approximation errors and, consequently, require a more conservative bound on $\Psi\Psi^\top$. As a consequence, the bound $\gamma \gamma^\top$ in Assumption~\ref{assump:bound2} may need to be chosen larger, which directly affects the feasibility of condition~\eqref{eq:thm} through matrix $\mathcal{Y}$. Lastly, since the solution of condition~\eqref{eq:thm} is generally not unique, any feasible solution, through its components $\mathbfcal K$ and $\mathbfcal Q\succ0$, yields, with $\mathbfcal P=\mathbfcal Q^{-1}$, a valid virtual state-feedback control law certified by Theorem~\ref{thm:main}. Although different feasible solutions may result in different transient behaviors or control efforts, the optimization of such performance metrics, while important, lies beyond the scope of the present work. \hfill$\square$
	\end{remark}
	
	We emphasize that the matrix $\mathcal{S} \coloneq [\avec \;\; \mathcal{A}]$ is entirely unknown in~\eqref{eq:new-closed-form}. Importantly, we neither identify $\mathcal{S}$ at any stage nor require it in our proposed condition~\eqref{eq:thm}.  {Instead, through the use of the S-procedure, our design is valid for all $x_r$-dynamics consistent with the collected data and the assumed noise and disturbance bounds.}
	
	\begin{remark}\label{rem:noisy_data_}
		The data used in condition~\eqref{eq:thm} are collected from a single trajectory of the unknown \textsc{ct-PNS}. Hence, while the unknown components $\mathcal A$, $\avec$, $f$, and $b_n$ do not appear explicitly in~\eqref{eq:thm}, their effects are embedded in the collected data. In particular, $\mathcal A$ and $\avec$ enter the derivative data through~\eqref{eq:thm1_tmp4}, while $f$, $b_n$, the data-collection input, and the matched disturbance $d_n$ affect the data indirectly through the evolution of $x_n$ and, consequently, through $\Oi_2$ and the resulting evolution of $x_r$. Notice also that, while we assume the elements of $\Oi_1$ and $\Oi_2$ are measured precisely, the system is subject to both unmatched and matched disturbances during data collection; the effect of the unmatched disturbance appears directly in condition~\eqref{eq:thm} through matrix $\mathcal{Y}$. The data matrix $\mathbfcal O_2$, which contains samples of $x_n$, is influenced by both the noisy input and the matched disturbance data $\D_2$; it enters condition~\eqref{eq:thm} through $\mathbfcal G$. The derivative data matrix $\Opoo$ is likewise noise-corrupted due to the approximation error in~\eqref{eq:FD}, whose effect is also reflected in~\eqref{eq:thm} through $\mathcal{Y}$. Finally, observe that $\In$ does not explicitly appear in~\eqref{eq:thm} because $x_n$ is treated as the virtual input for the $x_r$-dynamics; this is also evident from~\eqref{eq:thm1_tmp4}. This absence enables our approach to handle noisy input data without requiring any additional assumptions.
		\hfill$\square$
	\end{remark}
	
	In the following subsection, we proceed to present our data-driven ASSOSM control framework.
	
	\subsection{ASSOSM Controller Design}\label{subsec:controller}
	Here, we describe the design of the ASSOSM controller based on the data-driven results established in Section~\ref{subsec:manifold}.  {To this end,} we first recall that according to~\eqref{eq:sliding_manifold} and Theorem~\ref{thm:main}, the proposed data-driven sliding variable is
	\begin{align}
		\sigma = x_n - \mathbfcal{K}(x_r)\mathbfcal{P} x_r. \label{eq:new-sliding-manifold}
	\end{align}
	 {Evidently,} the sliding variable $\sigma$ depends on the data-driven components $\mathbfcal{K}$ and $\mathbfcal{P}$, which are obtained according to Theorem~\ref{thm:main}.  {Consequently, one can deduce that} the ASSOSM controller  {to be designed} also depends on  {these data-driven components.}
	
	 {Before presenting the design of the ASSOSM controller, which guarantees the finite-time convergence of the sliding variable~\eqref{eq:new-sliding-manifold} and its time derivative to zero, two key remarks are in order.} First,  {standard SOSM control approaches, \emph{e.g.}, the work by~\citet{bartolini1998chattering}, typically require knowledge of certain bounds, \emph{e.g.}, an upper bound on a term appearing in the second time derivative of the sliding variable, which is neither feasible nor realistic in our data-driven setting.} By employing the ASSOSM control framework, we circumvent this classical requirement, making the approach more suitable for a data-driven context.  {Second,} the resulting control input is continuous, which offers a clear advantage in practical implementations.
	
	Interpreting~\eqref{eq:new-sliding-manifold} as the output of the \textsc{ct-PNS}~\eqref{eq:sys_reformulated}, the resulting input--output map has relative degree one\footnote{That is, the control input \( u \) appears explicitly in the first derivative of the sliding variable~\eqref{eq:new-sliding-manifold}, \emph{i.e.,} \( \dot{\sigma} \).}. However, designing the ASSOSM controller, which yields a continuous control input, requires artificially increasing the relative degree to two. To do so, we introduce the auxiliary variables $\varsigma_1 \coloneq \sigma$ and $\varsigma_2 \coloneq \dot{\sigma}$,  {and construct} the auxiliary system
	\begin{align}
		\begin{split}
			\dot{\varsigma}_1 = \varsigma_2, \qquad \dot{\varsigma}_2 = \Delta + \Lambda \nu, \qquad  {\dot{u} = \nu,}
		\end{split}\label{eq:aux-sys}
	\end{align}
	where $\Lambda = b_n$, $\Delta = \ddot\sigma-b_n\nu$\footnote{The expression for $\ddot\sigma$ is rather long and is omitted to save space.}, and $\nu$ denotes the discontinuous control input that is to be designed to ensure that $\varsigma_1$ and $\varsigma_2$ reach zero in finite time.
	
	 {Notice that} $\varsigma_2$ is unmeasurable, as it depends on  {the matched and unmatched disturbances $d_n$ and $d_r$.} However, to design $\nu$, and, in particular, to design the adaptation law, the value of $\varsigma_2$ at each time instant is required. Thus, we employ Levant's differentiator~\citep{levant1998robust,levant2003higher} to estimate $\varsigma_2$ with high accuracy; in theory, the estimate becomes exact after a finite time. More concretely, to estimate $\varsigma_2$ by using Levant's differentiator, we have
	\begin{subequations}\label{eq:Levant}
		\begin{align}
			\dot{\hat{\varsigma}}_1 &= - \mu_0 \left| \hat{\varsigma}_1 - \varsigma_1 \right|^{\frac{1}{2}} \; \mathrm{sign}(\hat{\varsigma}_1 - \varsigma_1) + \hat{\varsigma}_2,\\
			\dot{\hat{\varsigma}}_2 &= -\mu_1 \; \mathrm{sign}(\hat{\varsigma}_1 - \varsigma_1),
		\end{align}
	\end{subequations}
	where $\hat{\varsigma}_1$ and $\hat{\varsigma}_2$ denote the estimates of $\varsigma_1$ and $\varsigma_2$, respectively. A possible choice of the differentiator parameters is $\mu_0 = 1.5\, \mathcal{L}^{\frac{1}{2}}$ and $\mu_1 = 1.1\, \mathcal{L}$, where $\mathcal{L} \in \Rp$  {is assumed to be chosen such that the differentiator can converge.}
	It is important to allow Levant's differentiator~\eqref{eq:Levant} sufficient time to converge and obtain a reliable estimate of $\varsigma_2$. Notably, the differentiator is proven to converge in finite time.
	
	Having adopted the Levant's differentiator to estimate $\varsigma_2$, we can now design the following discontinuous control input $\nu$ to drive both $\varsigma_1$ and $\varsigma_2$ to zero in finite time~\citep{incremona2016adaptive}:
	\begin{subequations}\label{eq:adaptive_SSOSM}
		\begin{align}
			\nu = - \Upsilon_{\!\mathrm{ad}} \; \mathrm{sign} \Big(\varsigma_1 - \frac{1}{2} \varsigma_1^{\max}\Big)\! , \label{eq:nu}
		\end{align}
		where $\Upsilon_{\!\mathrm{ad}}$ is the adaptive control amplitude to be designed, and $\varsigma_1^{\max}$ denotes the extremal value of $\varsigma_1$  {along each parabolic arc of the trajectory.} To design $\Upsilon_{\!\mathrm{ad}}$, we have
		\begin{align}
			\dot \Upsilon_{\!\mathrm{ad}} = \begin{cases}
				\eta_1 \vert \varsigma_1 \vert + \eta_2  \vert \hat{\varsigma}_2 \vert, & \text{if } \vert \varsigma_1 \vert > \vert  \Theta  \vert,\\
				0, & \text{otherwise,}
			\end{cases}\label{eq:adaptive_gain}
		\end{align}
	\end{subequations}
	where $\eta_1, \eta_2 \in \Rp$ are design parameters, $\Theta$ denotes  {the maximum of the sequence of the values of} $\varsigma_1$ stored as $\varsigma_1^{\max}$, and $\Upsilon_{\!\mathrm{ad}}(\mathsf{t}_0) = \Upsilon_{\!\mathrm{ad}_0}$. We note that,  {as the input $\nu$ is piecewise constant,} the actual control input $u(t) = u(\mathsf{t}_0)+ \int_{\mathsf{t}_0}^{t} \nu(s) \mathrm{d}s$  {is} continuous.
	
	\begin{remark}\label{rem:extremal}
		Levant's differentiator~\eqref{eq:Levant} can be employed to compute $\varsigma_1^{\max}$ by storing the value of $\varsigma_1$ at time instants when the sign of $\hat{\varsigma}_2$ changes (cf. Fig.~\ref{fig:Levant}).
		\hfill$\square$
	\end{remark}
	
	\begin{remark}\label{rem:levant_state_derivative}
		We emphasize that Levant's differentiator is not used to approximate the state derivative data matrices in~\eqref{eq:DE-DATA}. Indeed, the approximated state derivative data matrix $\Opoo$ appearing in condition~\eqref{eq:thm} is obtained exclusively via~\eqref{eq:FD}. Levant's differentiator is introduced only at the ASSOSM control-design stage, after the sliding variable has been constructed. Specifically, it is employed to estimate $\varsigma_2$ from $\varsigma_1$, which is required in~\eqref{eq:adaptive_gain}, and to detect $\varsigma_1^{\max}$, which is used in~\eqref{eq:nu}. \hfill$\square$
	\end{remark}

	\begin{figure}[t!]
		\resizebox{1.022\linewidth}{!}{
			\begin{tikzpicture}[->,auto,node distance=3cm,scale=1]
				\tikzset{point/.style={coordinate},
					block/.style ={draw, thick, rectangle, minimum height=4em, minimum width=6em},
					line/.style ={draw, very thick,-},
				}
				\node[block, align=center,draw=blue,fill=gray!5] (a) {Levant’s\\differentiator\\\eqref{eq:Levant}};
				\node (in)    (in1)    [left = 0.7 cm  of a]{$\varsigma_1$};
				\node[block, align=center,draw=blue,fill=gray!5] (b) [right = 1.5 cm  of a]{Zero crossing\\test};
				\node[block, align=center,draw=blue,fill=gray!5] (c) [right = 1.5 cm  of b]{Storing\\ $\varsigma_1$};
				\node (out)    (out1)    [right = 0.7 cm  of c]{$\varsigma_1^{\max}$};
				\draw[->] (in1)   -- (a);
				\draw[->] (c)   -- (out1);
				\draw[->] (a) -- (b) node[midway,above]{$\hat\varsigma_2$};
				\draw[->] 
				(b.east) -- ++(0.2,0) 
				to[spst] 
				++(1,0) coordinate (sw)
				-- (c.west)
				node[pos=-1.5, above, yshift=10pt] {if true};
			\end{tikzpicture}
		}
		\caption{Schematic of using Levant’s differentiator~\eqref{eq:Levant} to compute $\varsigma_1^{\max}$.}
		\label{fig:Levant}
	\end{figure}
	
	Note that in the non-adaptive version~\citep{bartolini1998chattering}, the control amplitude $\Upsilon_{\!\mathrm{ad}}$ is constant and should satisfy
	\begin{align}
		\Upsilon_{\!\mathrm{ad}} = \Upsilon_{\!\mathrm{ad}_0}  > \max   \Big\{\frac{\bar{\Delta}}{\underline{\Lambda}}, \frac{4 \bar{\Delta}}{3 \underline{\Lambda} -  \bar{\Lambda}}\Big\}\!,\label{eq:bart}
	\end{align}
	where $\bar{\Delta}, \, \underline{\Lambda},$ and $\bar{\Lambda}$ are \emph{known} positive constants satisfying
	\begin{align}
		\vert \Delta \vert \leq \bar{\Delta}, \quad 0<\underline{\Lambda} \leq \Lambda \leq \bar{\Lambda}. \label{eq:assump3}
	\end{align}
	However, finding such bounds is nearly impossible in our data-driven setting, thus demonstrating the advantage of adopting the ASSOSM control framework.
	
	\begin{remark}\label{rem:S-GAS}
		Note that the control amplitude $\Upsilon_{\!\mathrm{ad}}$ in~\eqref{eq:adaptive_gain} increases adaptively when the magnitude of the sliding variable tends to exceed that of $\Theta$; otherwise, $\Upsilon_{\!\mathrm{ad}}$ retains its previous value. As shown by~\citet{incremona2016adaptive}, if $\Upsilon_{\!\mathrm{ad}}$ is updated according to~\eqref{eq:adaptive_gain}, it satisfies~\eqref{eq:bart} in finite time. Consequently, as demonstrated by~\citet{bartolini1998chattering}, $\varsigma_1$ and $\varsigma_2$ reach zero in finite time.
		\hfill$\square$
	\end{remark}
	
	The following theorem establishes that, under the ASSOSM control input \(u\), the state trajectories of the closed-loop system are S-GUB.
	
	\begin{theorem}\label{coro}
			Let Assumptions~\ref{assump:bound}--\ref{assump:bound2} hold,  {and suppose that condition~\eqref{eq:thm} in Theorem~\ref{thm:main} is feasible.} Consider the auxiliary system~\eqref{eq:aux-sys} under the discontinuous control input \(\nu\) in~\eqref{eq:nu}, with the adaptive control amplitude \(\Upsilon_{\!\mathrm{ad}}\) given by~\eqref{eq:adaptive_gain}, together with Levant's differentiator~\eqref{eq:Levant}. Assume that \(\mathsf{t}_0 \geq t_{\mathcal{L}}\), with \(t_{\mathcal{L}}\) being the finite time required for the convergence of the differentiator.
			Assume further that, for every prescribed bounded set of initial conditions \(\mathcal X_0\subset\R^n\), the corresponding reaching phase is well defined and there exists an unknown finite constant \(\bar\Delta_{\mathcal X_0}\in\Rp\) such that \(|\Delta(t)|\leq \bar\Delta_{\mathcal X_0}\) along this reaching phase for all \(x(\mathsf{t}_0)\in\mathcal X_0\) and all admissible disturbances satisfying Assumption~\ref{assump:bound}.
			Then, for each such set $\mathcal X_0$,
			within a finite time \(t_r \geq t_d \geq \mathsf{t}_0\), where \(t_d\) is the time instant at which~\eqref{eq:bart} holds, the auxiliary system state variables \(\varsigma_1\) and \(\varsigma_2\) are driven to the origin of the auxiliary system state space. That is,  {for all $t \geq t_r$,}  a sliding mode on the manifold $\sigma  {(t)} = 0$, with $\sigma$  {as} in~\eqref{eq:new-sliding-manifold}, is enforced, implying that $x_n (t) = \mathbfcal{K}(x_r (t))\mathbfcal{P} x_r(t)$.
			Consequently, the state trajectories \(x\) of the closed-loop \textsc{ct-PNS}~\eqref{eq:sys_reformulated}, under the continuous control input \(u\) obtained by \(\dot u=\nu\), are S-GUB:
			\begin{align}
				\limsup_{t\to\infty}\|x(t)\| \leq \sqrt{ R_r^2+ R_n^2  }, \label{eq:UBound}
				\end{align}
			where $R_r \! \coloneq \! \sqrt{\frac{1}{\kappa \epsilon\lambda_{\min}(\mathbfcal P)}}\,\bar d_1$ and $R_n \! \coloneq \! \sup_{\|x_r\|\leq R_r} \vert \varphi(x_r)\vert$. Moreover, if \(d_r\equiv \mathbf{0}_{n-1}\), then the closed-loop system is S-GAS at $x = \mathbf{0}_n$.
			\hfill$\square$
			\end{theorem}
			
			\begin{proof}
				Since we employ Levant's differentiator and assume \(\mathsf{t}_0 \geq t_\mathcal{L}\), \(\varsigma_2\) is known exactly for all \(t\geq \mathsf{t}_0\), as the estimate \(\hat{\varsigma}_2\) becomes \emph{theoretically} exact after the finite convergence time \(t_\mathcal{L}\). This implies that the extremal value \(\varsigma_1^{\max}\) can, in principle, be detected with ideal accuracy using the procedure outlined in Remark~\ref{rem:extremal}.
				
				Moreover, by assumption, we know that for a prescribed bounded set of initial conditions $\mathcal X_0\subset\R^n$, there exists an unknown finite constant $\bar\Delta_{\mathcal X_0}\in\Rp$ such that $|\Delta(t)|\leq \bar\Delta_{\mathcal X_0}$ along the corresponding reaching phase for all $x(\mathsf{t}_0)\in\mathcal X_0$ and all admissible disturbances satisfying Assumption~\ref{assump:bound}. Hence, the gain condition~\eqref{eq:bart} is well defined for this prescribed set. As outlined in Remark~\ref{rem:S-GAS}, the adaptive gain $\Upsilon_{\!\mathrm{ad}}$ reaches, in finite time $t_d$, a value satisfying~\eqref{eq:bart}. Therefore, according to~\citet{incremona2016adaptive} and~\citet{bartolini1998chattering}, there exists a finite time $t_r\geq t_d\geq\mathsf{t}_0$ such that $\varsigma_1(t) = \varsigma_2(t)=0$ for all $t\geq t_r$. Since $\varsigma_1=\sigma$ and $\varsigma_2=\dot\sigma$, the sliding manifold $\sigma=0$ is reached in finite time, and the system state is maintained thereon for all $t \geq t_r$. Thus, from~\eqref{eq:new-sliding-manifold}, we have $x_n(t)=\mathbfcal K(x_r(t))\mathbfcal P x_r(t)$ for all $t\geq t_r$. Since \(|\Delta(t)|\leq\bar\Delta_{\mathcal X_0}\) during the reaching phase, the ASSOSM convergence result established earlier remains applicable. Consequently, \(\sigma=\varsigma_1\) remains bounded on the finite reaching interval \([\mathsf t_0,t_r)\).
				
				We now show that no divergence occurs during the reaching phase, \emph{i.e.}, when $t\in[\mathsf{t}_0,t_r)$ and $\sigma(t) \neq 0$. From~\eqref{eq:new-sliding-manifold}, one has $x_n=\varphi(x_r)+\sigma$. Hence, inspecting~\eqref{eq:sys_reformulated}, we have
					\[
					\dot x_r = \mathcal A\Omega(x_r)x_r+\avec\varphi(x_r)+d_r+\avec\sigma .
					\]
					By Theorem~\ref{thm:main}, for the $x_r$-dynamics with $x_n=\varphi(x_r)$ (\emph{i.e.}, $\sigma = 0$), the function $\V(x_r)=x_r^\top\mathbfcal P x_r$ satisfies
					\begin{align}
						\dot{\V}(x_r) \leq -\kappa \V(x_r)+\rho\|d_r\|^2 . \label{eq:temp1}
					\end{align}
					Therefore, along the off-manifold dynamics (\emph{i.e.}, $\sigma \neq 0$), one obtains
					\[
					\dot{\V}(x_r) \leq -\kappa \V(x_r) +\rho\|d_r\|^2 +2x_r^\top\mathbfcal P\avec\sigma .
					\]
					Using Cauchy--Schwarz and Young's inequalities, one has
					\[
					2x_r^\top\mathbfcal P\avec\sigma \leq \frac{\kappa}{2}\V(x_r) + \frac{2}{\kappa}\|\sqrt{\mathbfcal P}\avec\|^2 \sigma^2 .
					\]
					Therefore, during the reaching phase, we have
					\[
					\dot{\V}(x_r) \leq -\frac{\kappa}{2}\V(x_r) +\rho\|d_r\|^2 + \frac{2}{\kappa}\|\sqrt{\mathbfcal P}\avec\|^2 \sigma^2 .
					\]
					Consequently, since \(d_r\) is bounded under Assumption~\ref{assump:bound} and \(\sigma\) is
					bounded during the reaching phase by the ASSOSM convergence result invoked
					above, \(x_r\) remains bounded during this phase. As \(\varphi\) is a
					polynomial function in $x_r$, it remains bounded on this interval. Hence, \(x_n=\varphi(x_r)+\sigma\) remains bounded during this phase. Thus, no finite escape occurs during the reaching
					phase.
				
				For all $t\geq t_r$, one has $\sigma(t)=0$, and therefore~\eqref{eq:temp1} holds, which, considering Theorem~\ref{thm:main}, straightforwardly implies that
				\[
				\limsup_{t\to\infty}\|x_r(t)\| \leq R_r \coloneq \sqrt{\frac{\rho}{\kappa\underline{\alpha}}}\,\bar d_1 = \sqrt{\frac{1}{\kappa \epsilon\lambda_{\min}(\mathbfcal P)}}\,\bar d_1.
				\]
				Additionally, for all \(t\geq t_r\), $x_n(t)=\varphi(x_r(t))$, implying that
				\[
				\limsup_{t\to\infty}\vert x_n(t)\vert \leq R_n \coloneq \sup_{\|x_r\|\leq R_r}\vert\varphi(x_r)\vert.
				\]
				Thus, upon the definition of the Euclidean norm, \eqref{eq:UBound} evidently holds. It also follows that, if \(d_r\equiv \mathbf{0}_{n-1}\), the closed-loop system is S-GAS at $x = \mathbf{0}_n$, concluding the proof.
			\end{proof}
			
			Observe that once the data matrices are constructed, condition~\eqref{eq:thm} is solved as a feasibility problem. If feasible, any solution returned provides valid $\mathbfcal K$ and $\mathbfcal Q\succ0$, from which $\mathbfcal P=\mathbfcal Q^{-1}$ is computed. The sliding variable in~\eqref{eq:new-sliding-manifold} is then constructed, and the ASSOSM controller is designed accordingly. Thus, condition~\eqref{eq:thm} provides a computational step required for the proposed data-driven ASSOSM control design. Algorithm~\ref{Alg:1} summarizes these steps.
			
			\begin{algorithm}[t!]
				\caption{Data-driven design of ASSOSM control}\label{Alg:1}
				\begin{center}
					\begin{algorithmic}[1]
						\REQUIRE 
						Assumptions~\ref{assump:bound}--\ref{assump:bound2}, $\bar{\psi}$, $\mathcal{M}$, and $\kappa, \varkappa \! \in \! \Rp$
						\STATE 
						Collect $\Oi_1$  {and} $\Oi_2$,  {construct} $\Opoo$  {using~\eqref{eq:FD},} and form $\mathbfcal{J}$ as in~\eqref{eq:dictionary_data}
						\STATE
						Compute $\mathds{D}$ and $\gamma \gamma^\top$ according to~\eqref{eq:bound_dr} and Remark~\ref{rem:assum2}
						\STATE
						Fix the degrees of $\mathbfcal{K}$ and $\varpi$, and solve~\eqref{eq:thm} using \textsf{SOSTOOLS}~\citep{prajna2004sostools} to obtain $\mathbfcal{Q}$, $\epsilon$, and the coefficients of $\mathbfcal{K}$ and $\varpi$
						\STATE
						Compute $\mathbfcal{P}$ as $\mathbfcal{Q}^{-1}$  {and obtain} $\varphi(x_r)=\mathbfcal{K}(x_r)\mathbfcal{P}x_r$
						\STATE
						Construct the sliding variable $\sigma$ as in~\eqref{eq:sliding_manifold}
						\STATE
						Employ Levant's differentiator~\eqref{eq:Levant} to estimate $\varsigma_2$
						\STATE
						Design the adaptive $\Upsilon_{\!\mathrm{ad}}$ as in~\eqref{eq:adaptive_gain}
						\STATE
						Compute $\varsigma_1^{\max}$ according to Fig.~\ref{fig:Levant}
						\STATE
						Design the discontinuous input $\nu$ as in~\eqref{eq:nu}
						\STATE
						Compute the continuous control input $\dot u = \nu$
						\ENSURE
						Semiglobal ultimate boundedness of $x$ as in~\eqref{eq:UBound}, matched disturbance rejection
					\end{algorithmic}
				\end{center}
			\end{algorithm}
			
			\section{Simulation Results}\label{sec:simul}
			In this section, we demonstrate the effectiveness of the proposed data-driven framework through a \textsc{ct-PNS}, described by
				\begin{align}
					\dot{x}_1 & = 0.5 x_1^2 + x_2 + d_1, \quad \dot{x}_2 = f(x) + u + d_2, \label{eq:sys_example}
				\end{align}
				where $d_1(t) = 0.2 \sin(t)$, $d_2(t) = \tanh(2\sin(t))$, and $f(x) = \frac{x_1 x_2}{1 + x_1^2 + x_2^2} + \sin(x_1^2 + x_1 x_2)$. As the maximum degree of the polynomial function appearing in $\dot{x}_1$ is two, according to~\eqref{eq:sys_reformulated1} and~\eqref{eq:sys_reformulated}, we have $\mathcal{M}(x_1) = [x_1 ~~ x_1^2]^\top$ and $\Omega(x_1) = [1 ~~ x_1]^\top$. Moreover, we note that~\eqref{eq:sys_example} conforms to the structure of the \textsc{ct-PNS}~\eqref{eq:sys_reformulated} with $\mathcal{A} = [0 ~~ 0.5]$, $\avec = 1$, $b_2 = 1$, and the aforementioned $f(x)$, all of which are assumed to be fully unknown.
			
\begin{figure}[t!]
				\centering
				
				\subfloat[Nonlinear sliding variable\label{fig:states_EX1}]{
					\includegraphics[width=0.475\linewidth]{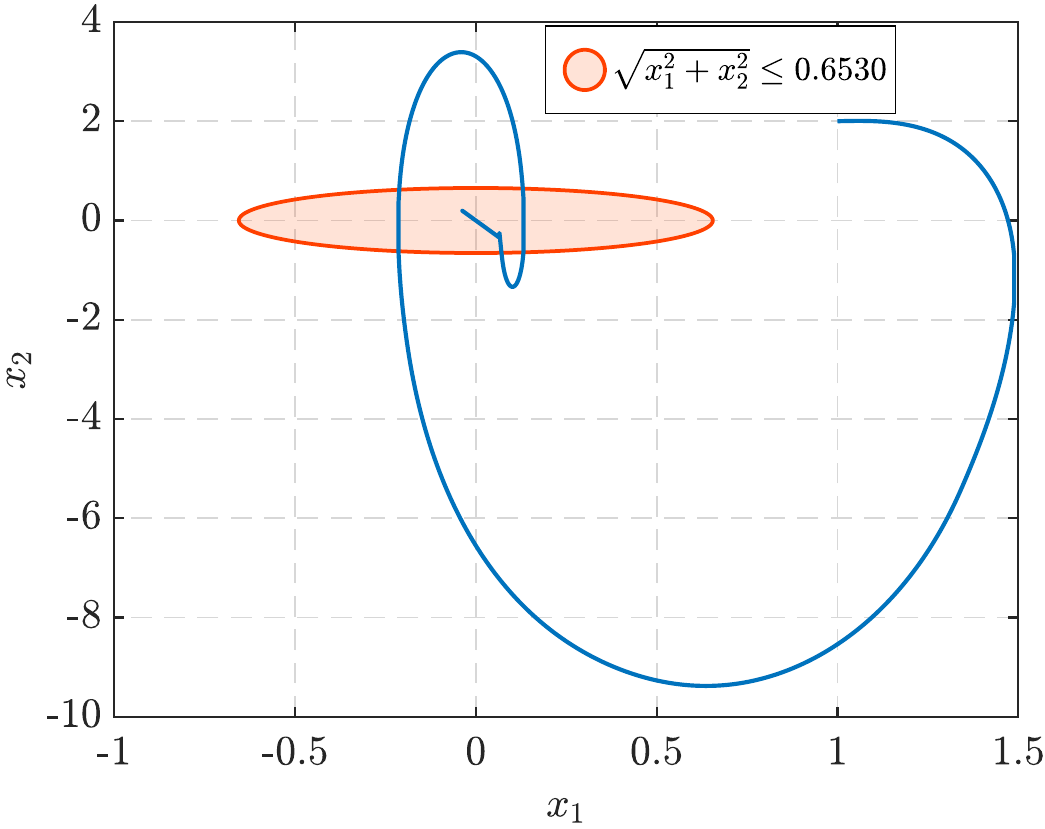}}\hspace{0.05cm}
				\subfloat[Linear sliding variable\label{fig:states_EX2}]{
					\includegraphics[width=0.475\linewidth]{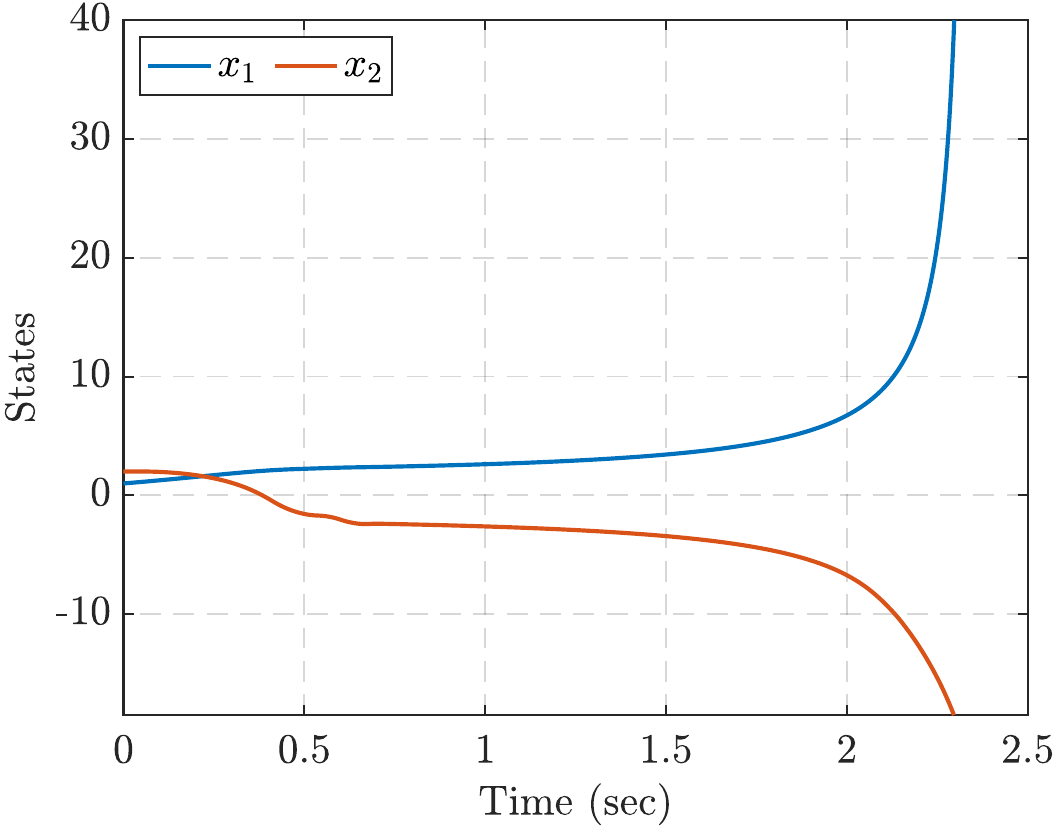}}
					\\
					\subfloat[Auxiliary system trajectory\label{fig:states_EX3}]{
						\includegraphics[width=0.475\linewidth]{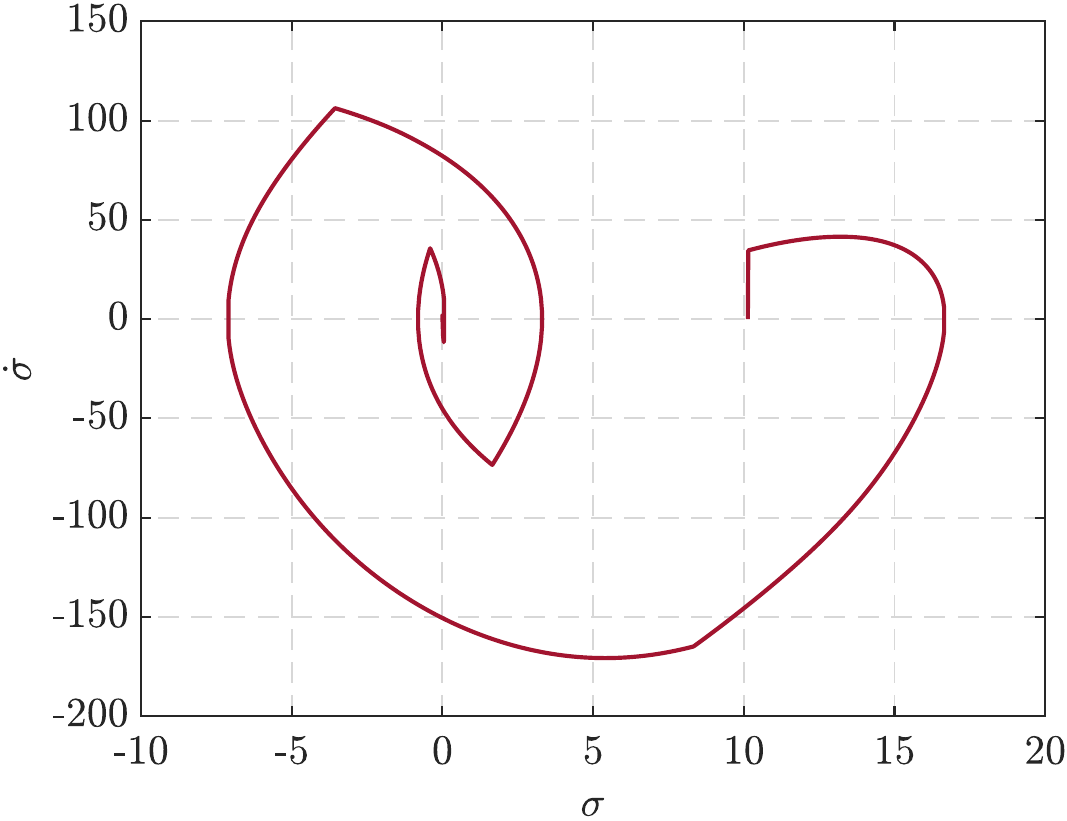}}\hspace{0.05cm}
					\subfloat[Levant's differentiator performance\label{fig:states_EX4}]{
						\includegraphics[width=0.45\linewidth]{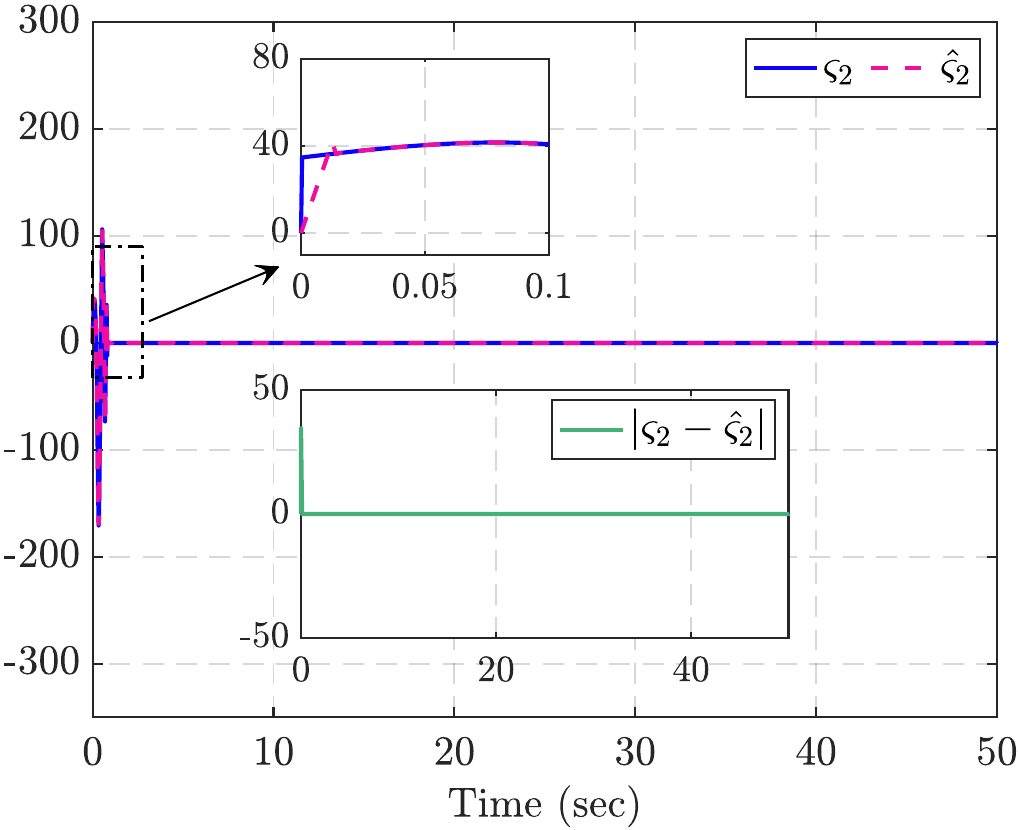}}
				
				\caption{(a) The phase portrait of the system, which demonstrates that, under the proposed data-driven design with the obtained \emph{nonlinear} sliding variable, the state trajectories of the closed-loop system are S-GUB, in accordance with Theorem~\ref{coro}. (b) The ineffectiveness of a linear sliding variable under the same simulation parameters. (c) The phase portrait of the auxiliary system~\eqref{eq:aux-sys}, demonstrating that $\sigma$ and $\dot{\sigma}$ are steered to zero. (d) The estimation of $\varsigma_2$ using Levant's differentiator. As can be observed, after a finite time, the estimate becomes sufficiently accurate.}
				\label{fig:EX1}
			\end{figure}

			To proceed with our approach, we first collect noisy data by applying the arbitrary input $u(t) = -1.5 x_2(t) -0.7 x_1(t) + 0.7 \sin(3t) + 0.4 \cos(5t)$ and setting the initial condition to $x(0) = [1.84 ~ - \! 2.2425]^{\top}$. Note that this choice of control input is solely made for the sake of data collection. The sampling time is set to $\tau = 0.01$, and we gather $\mathcal{T} = 120$ samples. Under Assumption~\ref{assump:bound}, we consider $\bar{d}_1 = 0.25$, a value that exceeds the actual bound of $0.2$. Thus, according to~\eqref{eq:bound_dr}, we have $\mathds{D} = 7.5$. Furthermore, we note that the collected data satisfy Assumption~\ref{assump:bound2} with $\gamma = 0.5467$. Having completed the first two steps of Algorithm~\ref{Alg:1}, we fix $\kappa = 2.5$ and $\varkappa = 0.5$, set the degrees of $\mathbfcal{K}$ and $\varpi$ to two, and solve~\eqref{eq:thm}, which yields\footnote{All values reported herein are computed using full floating-point precision; however, the values presented are rounded to four decimal places for brevity.}
				\begin{align*}
					& \mathbfcal{K}(x_1) = -0.0434x_1^2 - 0.0036x_1 - 0.0866, \; \mathbfcal{P} = 60.8563,\\
					& \varpi(x_1)  = 0.0016x_1^2 + 0.0006x_1 + 0.0020, \; \epsilon = 0.0284,\\
					& \varphi (x_1) = -2.6434x_1^3 - 0.2205x_1^2 - 5.2708x_1,\\
					& R_r  = 0.1203, \; R_n = 0.6418 \overset{\eqref{eq:UBound}}{\Longrightarrow} \limsup_{t\to\infty}\|x(t)\| \leq 0.6530.
				\end{align*}
				The sliding variable $\sigma$ is also obtained as in~\eqref{eq:sliding_manifold}. Following the procedure described in Section~\ref{subsec:controller}, we now design the ASSOSM controller. The design parameter of Levant's differentiator~\eqref{eq:Levant} is selected as $\mathcal{L} = 3000$. For the adaptive control amplitude $\Upsilon_{\!\mathrm{ad}}$, defined in~\eqref{eq:adaptive_gain}, we set $\Upsilon_{\!\mathrm{ad}_0} = 0$, $\eta_1 = 50$, and $\eta_2 = 25$. Moreover, $\varsigma_1^{\max}$ is determined according to the procedure illustrated in Fig.~\ref{fig:Levant}. The discontinuous control input $\nu$ is obtained as in~\eqref{eq:nu}, and the continuous control input $u$ is computed accordingly.
				Simulation results for the unknown system~\eqref{eq:sys_example} under the designed ASSOSM control input $u$ are presented in Fig.~\ref{fig:EX1}. As illustrated, the results are fully consistent with the theoretical findings.
                
                Most notably, to demonstrate the advantage of the proposed \emph{nonlinear} sliding variable over standard linear choices, we also consider the linear sliding variable
				\(
				\sigma = x_2+x_1
				\)
				as an example. For this choice, the associated virtual linear feedback law \(\varphi(x_1)=-x_1\) renders the disturbance-free upper subsystem locally asymptotically stable at the origin, while a local ISS property can be established on suitable compact subsets contained in the region \(x_1<2\). However, this local property is not sufficient for the semiglobal objective considered in this paper. Indeed, using the same simulation parameters and the initial condition \(x(0)=[1~~2]^\top\), which satisfies \(x_1(0)<2\), Fig.~\ref{fig:states_EX2} shows that, during the reaching phase, the trajectory of \(x_1\) leaves the region \(x_1<2\), rendering the closed-loop response unstable.
				In contrast, when the proposed nonlinear sliding variable
				\(
				\sigma = x_2 +2.6434x_1^3 + 0.2205x_1^2 + 5.2708x_1
				\)
				is employed, the state trajectories remain bounded under the same simulation conditions. In particular, Fig.~\ref{fig:states_EX1} illustrates that the state trajectories of the closed-loop system are S-GUB, highlighting the effectiveness of the proposed approach.
			
			\section{Conclusion}
			This work developed a direct data-driven approach for designing ASSOSM controllers for a class of single-input nonlinear systems with partially unknown dynamics, subject to matched and unmatched disturbances. The system was decomposed into lower and upper subsystems. Noisy data from a finite-time experiment enabled the formulation of a data-dependent condition whose feasibility facilitated the design of a virtual state-feedback control law rendering the closed-loop upper dynamics ISS. Building on this result, we constructed a data-driven sliding variable to synthesize an ASSOSM controller for the full-order system, ensuring that the state trajectories are S-GUB. Extending this work to more general system models and HOSM controllers is a promising future research direction.

	\bibliographystyle{agsm}
	\bibliography{biblio}
	
	\newpage
	
	\begin{authorbio}[Behrad]{Behrad Samari} received his B.Sc. and M.Sc. degrees in electrical engineering, control major, from K. N. Toosi University of Technology, Tehran, Iran, and University of Tehran (UT), Tehran, Iran, in 2019 and 2022, respectively. He is currently working toward the Ph.D. degree in computer science at the School of Computing, Newcastle University, Newcastle upon Tyne, U.K. He is the Best Repeatability Prize Finalist at the 8$^{\text{th}}$ IFAC Conference on Analysis and Design of Hybrid Systems (ADHS), 2024. His research interests include system and control theory, data-driven approaches, and formal methods.
	\end{authorbio}
	
	\begin{authorbio}[Gian_Paolo]{Gian Paolo Incremona} is associate professor of automatic control at Politecnico di Milano. He was a student of the Almo Collegio Borromeo of Pavia, and of the Institute for Advanced Studies IUSS of Pavia. He received the bachelor’s and master’s degree’s summa cum laude in Electric Engineering, and the Ph.D. degree in Electronics, Electric and Computer Engineering from the University of Pavia in 2010, 2012 and 2016, respectively. From October to December 2014, he was with the Dynamics and Control Group at the Eindhoven Technology University, The Netherlands. He was a recipient of the 2018 Best Young Author Paper Award from the Italian Chapter of the IEEE Control Systems Society, and since 2018 he has been a member of the conference editorial boards of the IEEE Control System Society and of the European Control Association. At present, he is Associate Editor of the Journal Nonlinear Analysis: Hybrid Systems, International Journal of Control, and IEEE Control Systems Letters. His research is focused on sliding mode control, model predictive control and switched systems with application mainly to train control, robotics and power plants.
	\end{authorbio}
	
	\begin{authorbio}[Antonella]{Antonella Ferrara } received the M.Sc. degree in Electronic Engineering and the Ph.D. degree in Electronic Engineering and Computer Science from the University of Genoa, Italy, in 1987 and 1992, respectively. Since 2005, she has been Full Professor of Automatic Control at the University of Pavia, Italy. Her research activities are mainly in the area of nonlinear control, with a special emphasis on sliding mode control, and application to robotics, power systems and road traffic. She is author and co-author of more than 450 publications including more than 170 journal papers, 2 monographs (published by Springer Nature and SIAM, respectively) and one edited book (IET). She is currently serving as Associate Editor of Automatica, and Senior Editor of the IEEE Open Journal of Intelligent Transportation Systems. She served as Senior Editor of the IEEE Transactions on Intelligent Vehicles, as well as Associate Editor of the IEEE Transactions on Control Systems Technology, IEEE Transactions on Automatic Control, IEEE Control Systems Magazine and International Journal of Robust and Nonlinear Control. Antonella Ferrara is the Chair of the EUCA Conference Editorial Board, the Director of Operations of the IEEE Control Systems Society, the Vice-Chair for Industry of the IFAC TC on Nonlinear Control Systems (2024-2026), a member of the IFAC Industry Board and of the IFAC Conference Board.  Among several awards, she was a co-recipient of the 2020 IEEE Transactions on Control Systems Technology Outstanding Paper Award. She is a Fellow of IEEE, Fellow of IFAC, Fellow of the European Academy of Sciences (EurASc) and Fellow of AAIA. She is also a Senior Fellow of the Brussels Institute for Advanced Studies (BrIAS).
	\end{authorbio}
	
	\begin{authorbio}[Abolfazl]{Abolfazl Lavaei} is an Assistant Professor in the School of Computing at Newcastle University, United Kingdom. Between January 2021 and July 2022, he was a Postdoctoral Associate in the Institute for Dynamic Systems and Control at ETH Zurich, Switzerland. He was also a Postdoctoral Researcher in the Department of Computer Science at LMU Munich, Germany, between November 2019 and January 2021. He received the Ph.D. degree in Electrical Engineering from the Technical University of Munich (TUM), Germany, in 2019. He obtained the M.Sc. degree in Aerospace Engineering with specialization in Flight Dynamics and Control from the University of Tehran (UT), Iran, in 2014. He is the recipient of several international awards in the acknowledgment of his work including  Best Repeatability Prize (Finalist) at the ACM HSCC 2025, IFAC ADHS 2024, and IFAC ADHS 2021, HSCC Best Demo/Poster Awards 2022 and 2020, IFAC Young Author Award Finalist 2019, and Best Graduate Student Award 2014 at University of Tehran with the full GPA (20/20). His research interests revolve around the intersection of Control Theory, Formal Methods, and Statistical Learning Theory.
	\end{authorbio}

\end{document}